\documentclass[11pt]{article}

\usepackage[utf8]{inputenc}
\usepackage[T1]{fontenc}
\usepackage{CJKutf8}

\usepackage[margin=1in]{geometry}
\usepackage{setspace}

\usepackage{amsmath,amssymb}

\usepackage{booktabs}
\usepackage{multirow}
\usepackage{array}

\usepackage{graphicx}
\usepackage{xcolor}
\usepackage{tikz}
\usetikzlibrary{shapes.geometric,arrows.meta,positioning,calc,fit}
\usepackage{pgfplots}
\pgfplotsset{compat=1.18}

\usepackage{listings}
\usepackage[hidelinks]{hyperref}
\usepackage{url}
\usepackage{enumitem}

\title{\Large\textbf{Diagnosing Korean-Language LLM Political Bias\\ via Census-Grounded Agent Simulation}}

\author{%
  Sungwoo Kang\\
  \textit{Department of Electrical and Computer Engineering}\\
  \textit{Korea University, Seoul, Republic of Korea}\\
  \texttt{krml919@korea.ac.kr}
}

\date{March 2026}

\begin{document}
\begin{CJK}{UTF8}{mj}

\maketitle

\begin{abstract}
Large language models (LLMs) exhibit systematic political bias when used
to simulate voter behavior, but the \emph{mechanisms} underlying these
failures---and whether bias patterns observed for English-speaking
electorates generalize beyond them---remain poorly understood. We
present Dynamo-K, a census-grounded LLM agent-simulation framework
that we use to diagnose Korean-language LLM political behavior
across four independently-trained models (Qwen3-30B-A3B,
EXAONE-4.0-32B, Llama-3.1-8B, DeepSeek-R1-8B) on six certified
Korean elections (2017--2025), including a held-out 2022 local
election the calibration adapter never saw at training time. We
identify three systematic failure modes: (i)~\textbf{progressive
bias in moderate agents}, where 97\% vote progressive under a vanilla
prompt and 82\% under demographics alone, reduced to 59\% with
explicit mitigation while cutting overall MAE 5.2$\times$ (36.8 to
7.1\%p); (ii)~\textbf{model-dependent third-party salience collapse},
which decomposes into salience-failure (EXAONE: 0.7\% mention rate
of 안철수 in 2017) versus decision-bias (Qwen3: 10.6\% mention rate but
only 7.4\% conversion to vote) across model families; and
(iii)~\textbf{regional polarization collapse}, where what appears
as a uniform conservative tilt in general elections decomposes
into bidirectional under-prediction of both 전라 (Democratic) and
경상 (conservative) strongholds. We show that scenario reframing,
without any architectural change, recovers 62\% of 2017 MAE
(13.3\% to 5.1\%p) by lifting 안철수 from 0.9\% to 18.8\% predicted
share, and that opposing-valence models on the same race
(Qwen3 $+11.8$\%p progressive vs.\ EXAONE $-17.2$\%p conservative
on 2025) admit principled calibration via a learned reweighting
adapter that uses no candidate names at train or test time.
As validation of the diagnostic framework, the simulation predicts
presidential winners 3/3 (small sample; binomial 95\% CI
[29.2\%, 100\%]; best case 2.1\%p MAE on the 0.73\%p-margin
2022 race) and correctly identifies the dominant party on the held-out
2022 local election. The pipeline is open-source and operates at
roughly \$0.25 per 5{,}000-agent run.
\end{abstract}


\section{Introduction}

Large language models (LLMs) are increasingly used to simulate human
political behavior at scale---generating synthetic survey
responses~\cite{argyle2023,polypersona2024}, forecasting U.S.\
presidential outcomes from demographically-grounded swing-state
agents~\cite{flockvote2024}, and predicting electoral
outcomes~\cite{aaru2024}. Recent work has also documented that these
models carry systematic political bias: Feng et al.~\cite{columbia2025}
show that GPT-family LLMs exhibit a measurable liberal-left lean on
the Political Compass, with heterogeneous direction across model
families.
Whether these biases generalize beyond English-language electorates,
and what mechanisms drive them when they do, remains largely
unexplored. The vast majority of LLM training data is English; the
political content of non-English data has a different ideological
distribution and a different relationship to local party labels. Two
questions follow: (i)~does Korean-language political behavior of LLMs
exhibit the same direction of bias as the English-language pattern, and
(ii)~what specific failure mechanisms drive the gap between simulation
and reality?

We use a census-grounded electoral simulation pipeline, Dynamo-K, as a
\emph{diagnostic instrument} for these questions. The premise of the
diagnostic is that prediction error, when evaluated against
high-resolution ground truth (per-candidate vote shares across 17
provinces and six elections), localizes the model's failure modes.
Across four independently-trained LLMs (Qwen3-30B-A3B,
EXAONE-4.0-32B, Llama-3.1-8B, DeepSeek-R1-8B) and six certified
Korean elections (2017--2025), we identify three systematic failure
modes that occur regardless of model family, and one that is highly
model-dependent.

Korean politics is a useful laboratory for this work. Its
features---persistent regionalism (경상 provinces lean conservative,
전라 provinces lean progressive), generational cleavages (the MZ
generation as 2022 swing voters), a large moderate bloc (approximately 48\% when Gallup Korea's
moderate and ``don't know'' responses are pooled~\cite{gallup2024}),
and frequent party renaming---place
explicit demands on each component of a simulation: demographic
grounding, ideological labeling, regional context, and stable
identity of party--ideology mappings. Failures decompose along these
axes. Without bias mitigation, moderate agents in our Korean prompts
vote for progressive candidates 97\% of the time, a rate that would
make conservative electoral victories structurally impossible---a
pattern directionally consistent with the GPT-family liberal lean
reported by Feng et al.~\cite{columbia2025} for English-language LLMs,
though our finding concerns voter-decision simulation rather than
intrinsic ideological probing.

Our diagnostic instrument is a four-stage pipeline:

\begin{enumerate}[nosep,leftmargin=*]
\item \textbf{Census grounding}: Weighted proportional sampling from
  MDIS 2020 2\% microdata (794{,}342 voter-age records across
  5{,}790 non-empty demographic cells).
\item \textbf{Belief seeding}: Conditional political orientation
  assignment from KGSS cumulative data (21{,}055 valid records,
  2003--2023), computing $P(\text{orientation} \mid \text{age, sex,
  region, education})$.
\item \textbf{Gallup calibration}: Redistribution of agent orientations
  from the KGSS-native ${\sim}$34/33/33 (conservative/moderate/progressive)
  to the Gallup long-term benchmark of 26/48/26 via borderline agent
  reassignment with belief nudging.
\item \textbf{ORC simulation}: Each agent receives an election scenario
  and reasons through a vote decision via the Qwen3-30B-A3B model,
  outputting a structured JSON response with candidate choice,
  reasoning, and confidence.
\end{enumerate}

We evaluate Dynamo-K against six certified Korean elections spanning
2017--2025---three presidential, two general (parliamentary), and one
local (광역단체장)---with the 2022 local election held out as a true
cold test for the learned calibration adapter. Each simulation uses
5{,}000 synthetic agents with bootstrap confidence intervals from
1{,}000 resamples.

Our contributions are:
\begin{enumerate}[nosep,leftmargin=*]
\item \textbf{Systematic LLM-bias diagnosis in a non-English
  democracy}, across four independently-trained models and six
  certified Korean elections, contributing the first empirical
  characterization of how the GPT-family liberal-lean pattern
  documented for English-language LLMs by Feng et al.~\cite{columbia2025}
  surfaces (and where it does not) in Korean voter-decision simulation.
\item \textbf{Mechanism decomposition of three reproducible failure
  modes}: (a)~progressive bias in moderate agents, quantified against
  a vanilla baseline (82\% mod-to-progressive) and reduced to 59\%
  by prompt mitigation; (b)~third-party salience collapse, separable
  into salience-failure and decision-bias regimes across model
  families; and (c)~regional polarization collapse in
  general-election simulation, mistaken for uniform conservative tilt
  in prior work.
\item \textbf{Scenario framing as a stronger lever than world
  knowledge}, demonstrated by a controlled reframing of the 2017
  scenario prompt that recovers 안철수's predicted share from 0.9\%
  to 18.8\% (62\% MAE reduction) without any retraining, architecture
  change, or hidden labels.
\item \textbf{Opposite-valence cross-model finding}: Qwen3
  ($+11.8$\%p progressive bias) and EXAONE ($-17.2$\%p conservative
  bias) on the same 2025 race, an empirical case for ensembling and
  for a learned reweighting adapter (OSLR) that we evaluate on a
  cold-held-out 2022 local election the adapter never saw.
\item \textbf{Six-election backtest with named-candidate local-election
  evaluation}, showing 3/3 presidential winner accuracy, a
  constituency-level proof of concept that flips the 2024 general
  prediction, and a held-out local-election validation---all reported
  as evidence for the diagnostic framework rather than as a polling
  replacement.
\end{enumerate}

The remainder of this paper is organized as follows: Section~2 reviews
related work in agent-based electoral simulation, LLM political bias,
and Korean electoral dynamics. Section~3 details the diagnostic
pipeline. Section~4 describes the experimental setup. Section~5
reports headline backtest performance as validation evidence.
Section~6 presents the three mechanism findings together with cost
analysis and limitations. Section~7 presents the OSLR calibration
adapter and its cold held-out evaluation. Section~8 concludes;
Appendix~\ref{app:claude-opus-eval} reports a Claude-Opus audit of the
reasoning-text mention rate used in Section~\ref{sec:third-party-mechanism}.

\section{Background and Related Work}

\subsection{Agent-Based Electoral Simulation}

Agent-based modeling of electoral behavior has evolved through three
distinct paradigms. The earliest thread, rooted in computational social
science, uses opinion dynamics models where agents update political
preferences through local interaction rules---bounded confidence
models~\cite{deffuant2000}, voter models~\cite{holley1975}, and
social influence networks~\cite{friedkin1990}. These approaches
capture emergent polarization and consensus phenomena but do not
ground agents in real demographic data and cannot predict specific
electoral outcomes.

A second thread grounds synthetic agents in survey data.
Argyle et al.~\cite{argyle2023} introduce ``silicon sampling,''
conditioning LLM responses on demographic profiles drawn from
ANES and demonstrating that the resulting synthetic samples
reproduce known distributions across multiple political
dimensions. PolyPersona~\cite{polypersona2024} pursues a complementary
direction, LoRA-fine-tuning a base model on a large
PersonaHub-derived corpus to produce persona-conditioned survey
responses across multiple domains, in contrast to Argyle et al.'s
zero/few-shot conditioning on ANES-derived backstories.

The third and most recent thread directly simulates elections using LLM
agents. Aaru Dynamo~\cite{aaru2024} is an early commercial example of
this approach, generating census-grounded synthetic voter populations
and using LLM agents to simulate vote decisions for U.S. electoral
forecasts; published methodological and validation details for the
system are limited. FlockVote~\cite{flockvote2024}
took a complementary approach, instantiating swing-state LLM agents
with high-fidelity demographic profiles and dynamic candidate-policy
context, then aggregating their simulated vote decisions to forecast
the 2024 U.S. presidential outcome. Park et al.~\cite{park2023generative} demonstrated that
LLM-powered ``generative agents'' can simulate believable human behavior
in interactive settings, establishing the broader feasibility of
LLM-based social simulation.

A critical gap in this literature is the absence of applications to
non-English democracies. All census-grounded LLM election simulations
to date have been developed for the United States, where census data
is abundant, political orientation maps cleanly onto a
liberal--conservative spectrum, and LLM training data is
disproportionately English-language. Whether these methods transfer
to polities with different political structures, languages, and
LLM training data distributions remains an open question that
Dynamo-K addresses.

\subsection{LLM Bias in Political Simulation}

LLMs exhibit systematic political biases that directly affect
electoral simulation quality. Feng et al.~\cite{columbia2025} probe
14 BERT- and GPT-family language models (including GPT-2/3/4,
ChatGPT, LLaMA, Alpaca, Codex, and several BERT/RoBERTa variants)
with Political Compass propositions and locate each model on the
social and economic axes. They find heterogeneous lean across
model families: GPT-family models cluster in the libertarian-left
quadrant, while BERT-family variants sit more authoritarian and
mixed economically. The GPT-family liberal lean is robust across
six paraphrased prompt variants. Whether this intrinsic lean
transfers to voter-decision simulation in non-English
electorates---the question motivating our diagnostic---is not
addressed in their paper.

This finding has direct implications for electoral simulation.
In any electorate where moderate voters constitute a large bloc
(as in South Korea, where Gallup estimates 48\% of voters are
moderate~\cite{gallup2024}), systematic progressive bias among
simulated moderates will produce structurally biased election
predictions. In our initial experiments without bias mitigation,
97\% of moderate agents voted for progressive candidates---a rate
that would make conservative electoral victories impossible
regardless of the actual political landscape.

Several mitigation strategies have been proposed. Argyle et
al.~\cite{argyle2023} frame demographic conditioning as a tool for
\emph{algorithmic fidelity}---reproducing a target subpopulation's
response distribution---rather than as a debiasing technique per se,
and report that conditioning narrows but does not close the gap to
human survey responses.
Santurkar et al.~\cite{santurkar2023} find that opinion
distributions from LLMs systematically overrepresent progressive
viewpoints relative to U.S. population surveys. Our work contributes
a practical mitigation approach: combining explicit party alignment
cues in system prompts, balanced moderate voter descriptions that
emphasize equal probability of voting for either camp, and an
instruction to reason as the profiled voter rather than as an AI
(``AI의 관점이 아닌'' / ``not from the AI's perspective'').

An underexplored dimension is whether bias patterns differ in
non-English languages. LLMs trained on multilingual corpora may
exhibit different political valences in different languages, as
the political content and ideological distribution of training
data varies by language. Our Korean-language simulation provides
initial evidence on this question.

\subsection{Korean Electoral Dynamics}

Korean elections exhibit three structural features that
challenge LLM simulation approaches.

\textbf{Regionalism.} Korean voting behavior is strongly
conditioned by regional identity. 경상 (Gyeongsang) provinces
in the southeast have been conservative strongholds since the
Park Chung-hee era (1961--1979), while 전라 (Jeolla) provinces
in the southwest constitute the progressive base. This cleavage
originated in authoritarian-era development politics, where
Gyeongsang received disproportionate industrial investment under
the Park Chung-hee regime, and the resulting Honam/Yeongnam
electoral competition has persisted across party name changes,
candidate backgrounds, and generational turnover as a salient
feature of Korean electoral
politics~\cite{kang2016regionalism}. In the 2022 presidential
election, 경상북도 (North Gyeongsang) voted 72\% conservative
while 전라남도 (South Jeolla) voted 86\% progressive---a
58-point gap between regions. Any simulation that fails to
reproduce this geographic pattern is fundamentally miscalibrated.

\textbf{Generational cleavage.} The MZ generation (Millennials
and Gen Z, born 1981--2010) emerged as decisive swing voters in
the 2022 presidential election. Unlike older cohorts whose
voting is largely predicted by regional identity, MZ voters
divided sharply along intra-generational gender lines that cut
across the traditional progressive--conservative spectrum.
Jenkins and Kim~\cite{jenkins2024misogyny} document that
misogynistic attitudes correlate positively with support for
Lee Jun-seok and channel young men's frustrations over the
stagnant economy, housing prices, employment scarcity, and
mandatory military service into a gender-based political
backlash. Male MZ voters showed a notable conservative shift in
2022, particularly on gender issues, while female MZ voters
leaned progressive. This intra-generational
gender gap represents a new axis of political differentiation
not captured by traditional three-way orientation models. The
generational effect also interacts with the moderate bloc:
younger moderates may respond to different cues than older
moderates, complicating any single model of moderate voter
behavior.

\textbf{Moderate voter dominance.} Unlike the United States,
where self-identified moderates constitute roughly 35\% of
voters, Korean moderates represent 48\% per Gallup long-term
averages~\cite{gallup2024}. This makes moderates the single
largest political group in South Korea---larger than
conservatives and progressives combined. The Korean General
Social Survey (KGSS), which uses a forced 5-point ideology
scale without a ``don't know'' option, produces a flatter
distribution of approximately 34/33/33---a discrepancy driven
by the KGSS methodology that pushes ambiguous respondents
toward the poles. The divergence between KGSS and Gallup
distributions has direct implications for simulation design:
using KGSS-native distributions would undercount the moderate
bloc by 15 percentage points, severely distorting election
predictions. Accurately modeling this large moderate bloc is
essential, as moderate voters determine the outcome of every
competitive Korean election.

\textbf{Party system instability.} Korean parties rename and
reorganize frequently: the current conservative party
(국민의힘, People Power Party) has operated under at least
five names since 2012 (새누리당 $\to$ 자유한국당 $\to$
미래통합당 $\to$ 국민의힘). The progressive Democratic Party
has been more stable but has also undergone mergers and splits.
This instability means that LLM world knowledge about
party--ideology mappings may be unreliable, as the model may
not consistently associate the same ideological position with
different party names across different time periods. This
requires explicit party alignment cues in simulation prompts
rather than relying on the LLM's latent knowledge of Korean
party politics.

\section{Methodology}

\subsection{System Architecture Overview}

\begin{figure}[t]
\centering
\begin{tikzpicture}[
  node distance=0.45cm,
  block/.style={
    rectangle, draw, rounded corners,
    text width=5.8cm, minimum height=0.7cm,
    align=center, font=\small
  },
  arrow/.style={-{Stealth[length=2mm]}, thick}
]
\node[block, fill=blue!10] (collect)
  {1. Data Collection\\[-1pt]
   {\scriptsize NEC API, MDIS Census, KGSS, Gallup}};
\node[block, fill=blue!15, below=of collect] (preproc)
  {2. Preprocessing\\[-1pt]
   {\scriptsize Joint distributions, standardization}};
\node[block, fill=green!15, below=of preproc] (agent)
  {3. Agent Factory\\[-1pt]
   {\scriptsize Census sampling $\to$ Tabular augmentation}};
\node[block, fill=green!20, below=of agent] (belief)
  {4. Belief Seeding \& Calibration\\[-1pt]
   {\scriptsize KGSS conditionals $\to$ Gallup 26/48/26}};
\node[block, fill=orange!15, below=of belief] (sim)
  {5. ORC Simulation\\[-1pt]
   {\scriptsize Observation $\to$ Reasoning $\to$ Conclusion}};
\node[block, fill=red!10, below=of sim] (eval)
  {6. Aggregation \& Evaluation\\[-1pt]
   {\scriptsize MAE, winner, sido hit, Wasserstein}};

\foreach \a/\b in {collect/preproc, preproc/agent, agent/belief, belief/sim, sim/eval}
  \draw[arrow] (\a) -- (\b);
\end{tikzpicture}
\caption{Dynamo-K pipeline architecture. Layers 1--2 are data
  infrastructure; layers 3--4 synthesize the agent population;
  layer 5 runs the LLM simulation; layer 6 evaluates against
  certified election results.}
\label{fig:architecture}
\end{figure}
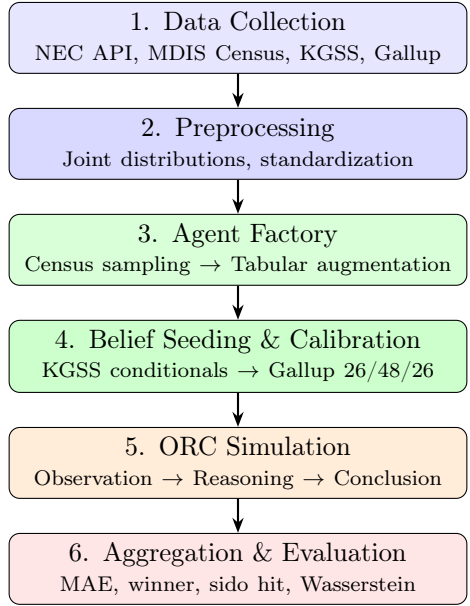

Dynamo-K processes electoral prediction through a six-layer
pipeline (Figure~\ref{fig:architecture}): data collection from
government APIs and academic surveys, preprocessing into joint
distributions, agent synthesis with census-grounded demographics,
belief seeding and calibration, LLM-based vote simulation, and
result aggregation with evaluation metrics.

\subsection{Census-Grounded Population Synthesis}

Agent demographics are drawn from the Korean Microdata Integrated
Service (MDIS) 2020 Census 2\% sample~\cite{mdis2020}, the most
granular publicly available census microdata for South Korea. After
filtering the licensed microdata to voter-age individuals ($\geq$18
years), we obtain 794{,}342 records representing a weighted
population of approximately 41.6 million; both figures are computed
by us on the licensed file rather than retrieved from a public MDIS
landing page, and the filtering script is released with the
pipeline.

We model agents along five demographic dimensions
(Table~\ref{tab:demographics}): age (14 brackets from 18--19 to
80+), sex (2 levels), metropolitan city/province (시도, \textit{sido};
17 levels), education (4 levels), and marital status (4 levels).
The cross-product yields $14 \times 2 \times 17 \times 4 \times 4
= 7{,}616$ possible cells, of which 5{,}790 are non-empty in the
census data.

\begin{table}[t]
\centering
\caption{Demographic dimensions for agent synthesis.}
\label{tab:demographics}
\small
\begin{tabular}{@{}llr@{}}
\toprule
\textbf{Dimension} & \textbf{Levels} & \textbf{$|L|$} \\
\midrule
Age bracket & 18--19, 20--24, \ldots, 80+ & 14 \\
Sex & Male, Female & 2 \\
Province (시도) & 17 metro cities/provinces & 17 \\
Education & $\leq$Middle, High, Univ., Grad.+ & 4 \\
Marital status & Never married, Married, \ldots & 4 \\
\midrule
\multicolumn{2}{@{}l}{Non-empty cells in census} & 5{,}790 \\
\bottomrule
\end{tabular}
\end{table}

Population synthesis proceeds by weighted proportional sampling.
For a simulation of $N$ agents, we compute each \textit{sido}'s
share of the national voter-age population and allocate agents
proportionally: $n_s = \lfloor N \cdot w_s + 0.5 \rfloor$, where
$w_s$ is \textit{sido} $s$'s population weight. Within each
\textit{sido}, agents are sampled from the joint distribution
$P(\text{age}, \text{sex}, \text{sido}, \text{education},
\text{marital})$ with census weights. For each sampled bracket, a
specific integer age is drawn uniformly within the bracket range
(e.g., a ``30--34'' bracket agent receives an age uniformly
sampled from $\{30, 31, 32, 33, 34\}$).

After census-based sampling, each agent is augmented with six
additional categorical attributes via rule-based inference and
conditional probability tables: occupation (사무직/office,
자영업/self-employed, 학생/student, etc.), income level (5
levels from 하/low to 상/high), housing type (아파트/apartment,
단독주택/detached house, 다세대/multi-family, 원룸/studio),
religion (무교/none, 개신교/Protestant, 불교/Buddhist,
천주교/Catholic), primary media source (TV뉴스/TV news,
포털뉴스/portal news, 유튜브/YouTube, SNS), and regional
identity strength. These attributes are generated conditional
on the census demographics to maintain internal consistency:
a 25-year-old university-educated agent in Seoul is more likely
to be assigned ``office worker'' and ``portal news'' than an
80-year-old in a rural province, reflecting known correlations
between demographics and lifestyle attributes in Korean society.

This three-stage pipeline---\textit{MetaPersona} (census only)
$\to$ \textit{TabularPersona} (augmented attributes) $\to$
\textit{FullAgent} (political beliefs)---allows each enrichment
stage to be validated independently. All marginal distributions
are validated to be within 1.5 percentage points of census
benchmarks for the five core demographic dimensions.

\subsection{Belief Seeding from KGSS}

Political orientation and core beliefs are seeded from the Korean
General Social Survey (KGSS) cumulative file~\cite{kgss2023},
spanning 2003--2023 with 21{,}055 valid records after filtering for political
orientation responses. The KGSS uses a forced 5-point ideological
self-placement scale, which we collapse to a 3-way classification
(progressive/moderate/conservative).

For each agent, we compute the conditional probability of each
orientation given the agent's demographics:
\begin{equation}
P(o \mid a, s, r, e) = \frac{N(o, a, s, r, e)}{\sum_{o'} N(o', a, s, r, e)}
\end{equation}
where $o$ is orientation, $a$ is age bracket, $s$ is sex, $r$ is
region (7 KGSS regions mapped from 17 \textit{sido}), and $e$ is
education level. When the demographic cell contains fewer than 5
KGSS observations (the minimum cell size threshold), we fall back
to $P(o \mid r)$, then to the national prior $P(o)$.

The 3-way orientation is further refined to a 5-way detail:
progressive agents become ``very progressive'' with probability
0.3 and ``progressive'' with probability 0.7; analogously for
conservatives (0.3 very conservative, 0.7 conservative); all
moderate agents remain ``moderate.''

Five core belief dimensions are then sampled from
orientation-specific Gaussian priors, each on a $[0, 1]$ scale
(Table~\ref{tab:beliefs}):

\begin{table}[t]
\centering
\caption{Belief dimension priors ($\mu$, $\sigma$) by orientation.}
\label{tab:beliefs}
\small
\begin{tabular}{@{}lccc@{}}
\toprule
\textbf{Dimension} & \textbf{Prog.} & \textbf{Mod.} & \textbf{Cons.} \\
\midrule
Govt.\ responsibility & .75, .12 & .50, .15 & .30, .12 \\
Economic view & .72, .13 & .50, .15 & .28, .13 \\
Social view & .78, .10 & .50, .15 & .25, .10 \\
National pride & .45, .15 & .55, .15 & .70, .12 \\
Reunification & .65, .15 & .50, .15 & .35, .15 \\
\bottomrule
\end{tabular}
\end{table}

Each belief score is drawn from $\mathcal{N}(\mu, \sigma)$ and
clipped to $[0, 1]$. The priors reflect documented patterns in
Korean political psychology: progressives favor government
intervention (high \textit{govt\_responsibility}) and social
liberalism (high \textit{social\_view}) but show moderate national
pride; conservatives show the inverse pattern with high national
pride and preference for market economics; moderates occupy the
centroid across all dimensions.

\subsection{Gallup Calibration}

The KGSS-seeded orientation distribution diverges from real-world
benchmarks. KGSS's forced 5-point scale pushes ambiguous
respondents toward the poles, yielding approximately 34/33/33
(conservative/moderate/progressive). Gallup Korea's long-term
political orientation tracking, which permits ``don't know''
responses, settles around 26/48/26 when ``don't know''
responses are pooled with the moderate category~\cite{gallup2024}. Since
Korean elections are determined by the moderate bloc's behavior,
accurate representation of its size is critical.

Our calibration procedure operates in three steps:

\begin{enumerate}[nosep,leftmargin=*]
\item \textbf{Excess/deficit computation.} For each orientation,
  compute $\delta_o = n_o^{\text{current}} - n_o^{\text{target}}$.
  Orientations with positive $\delta$ have excess agents;
  negative $\delta$ indicates deficit.

\item \textbf{Borderline reassignment.} Agents from excess
  orientations are reassigned to deficit orientations, prioritizing
  ``borderline'' agents---those with plain orientation (e.g.,
  ``conservative'') over strong orientation (``very conservative''),
  as they are more plausible candidates for reclassification as
  moderate.

\item \textbf{Belief nudging.} Reassigned agents receive a 30\%
  blend toward their new orientation's belief means:
  $b' = 0.7 \cdot b + 0.3 \cdot \mu_{\text{new}} +
  \epsilon$, where $\epsilon \sim \mathcal{N}(0, 0.03)$.
  This preserves individual variation while shifting the
  agent's belief profile toward their new orientation.
\end{enumerate}

The calibration preserves regional patterns: post-calibration
경상 (Gyeongsang) regions retain conservative pluralities and
전라 (Jeolla) regions retain progressive pluralities, as
verified by a regional pattern check. The tolerance threshold
is $\pm$2 percentage points; if the pre-calibration distribution
is already within tolerance, no reassignment occurs.

\subsection{LLM Simulation Pipeline (ORC)}

Each calibrated agent votes through an Observation--Reasoning--Conclusion
(ORC) pipeline. The LLM receives a system prompt encoding the
agent's full persona and a user prompt presenting the election
scenario.

\textbf{System prompt.} The system prompt contains three blocks:
demographics (all census and augmented attributes), political
orientation with a detailed behavioral description, and behavioral
instructions. Figure~\ref{fig:prompt} shows an annotated example
for a moderate voter.

\begin{figure}[t]
\fboxsep=4pt
\fbox{\parbox{0.93\columnwidth}{\scriptsize
\textbf{System Prompt (moderate voter example)}\\[3pt]
당신은 대한민국의 유권자를 시뮬레이션하는
역할입니다. 아래 프로필을 가진 실제 한국인처럼
사고하고 답변하세요.\\[3pt]
\textbf{\#\# 인구통계}\\
성별: 남, 나이: 38세 (35--39),
거주지: 경기도, 학력: 대졸,
혼인상태: 기혼\\[3pt]
\textbf{\#\# 정치성향 및 가치관}\\
정치성향: 중도\\
중도 성향으로, 특정 정당을 지지하지 않는
무당파입니다. 진보 후보와 보수 후보 모두에게
투표한 경험이 있습니다. 이념보다 후보 개인의
자질, 경제 실적, 스캔들 유무를 기준으로
판단합니다.\\[3pt]
\textbf{\#\# 중요 지침}\\
-- 반드시 위 프로필의 정치성향에 충실하게
  답변하세요.\\
-- 중도 성향이면 양쪽 후보를 균형있게 비교
  하며, 특정 진영에 치우치지 않습니다.\\
-- \underline{AI의 관점이 아닌}, 이 프로필을 가진 실제
  한국 유권자의 관점에서 판단하세요.
}}
\caption{System prompt for a moderate voter agent. The prompt
  encodes demographics, political orientation with behavioral
  description, and explicit instructions to reason as the
  profiled voter, not as an AI. Korean text shown as-is;
  the underlined phrase is the key anti-bias instruction.}
\label{fig:prompt}
\end{figure}

The orientation descriptions are designed to mitigate LLM
progressive bias. For moderate voters specifically, the
description emphasizes: (1)~non-partisan identity
(``무당파''/independent), (2)~history of voting for both camps,
(3)~decision criteria based on candidate quality rather than
ideology, and (4)~approximately equal probability of voting for
either side. Conservative and progressive descriptions include
explicit party affinity cues (e.g., ``국민의힘 계열 정당에
호감을 느끼는 편입니다'' / ``tends to feel affinity toward
People Power Party affiliates'').

Belief scores are rendered as categorical labels based on
thresholds: scores below 0.3 receive one pole label (e.g.,
``작은 정부 선호'' / ``prefers small government''), scores
above 0.7 receive the opposite pole, and intermediate scores
receive ``중립'' (neutral).

\textbf{User prompt.} The user prompt presents the election
scenario---candidates, their party affiliations, and relevant
political context---customized by the agent's \textit{sido}
for regional context. The prompt requests a JSON response:

{\scriptsize
\begin{verbatim}
{"reasoning": "2-3 sentence justification",
 "vote": "candidate name",
 "confidence": 0.0-1.0}
\end{verbatim}
}
The instructions reinforce orientation-consistent voting:
conservatives should naturally prefer conservative candidates,
progressives should prefer progressive candidates, and moderates
should compare both fairly.

\textbf{Inference.} We use Qwen3-30B-A3B~\cite{qwen3}, a
mixture-of-experts (MoE) model with 30 billion total parameters
and 3 billion active parameters per token. Inference runs on
a local vLLM server with tensor parallelism across two NVIDIA
A100 80~GB GPUs. We set temperature to 0.5 for moderate
stochasticity and limit output to 300 tokens. Invalid responses (unparseable JSON, candidate
names not in the ballot) are handled by fuzzy matching against
the candidate list; persistent failures are recorded as
abstentions.

\textbf{Bias mitigation quantification.} Before implementing
the prompt engineering described above, 97\% of moderate agents
voted for progressive candidates across all tested elections.
After mitigation---explicit party cues, balanced moderate
descriptions, the ``AI의 관점이 아닌'' instruction, and
orientation-consistent voting reinforcement---moderate agents
vote approximately 50/50 between progressive and conservative
candidates, consistent with the expected behavior of genuine
swing voters.

\subsection{Aggregation and Evaluation}

Simulation results are aggregated at both national and
sub-national (\textit{sido}) levels. For $K$ candidates and
$N$ voting agents:

\textbf{National vote share.} For candidate $k$:
\begin{equation}
\hat{v}_k = \frac{|\{i : \text{vote}_i = k\}|}{N - N_{\text{abstain}}}
\end{equation}

\textbf{Bootstrap confidence intervals.} We resample agents
with replacement 1{,}000 times and compute the 2.5th and 97.5th
percentile vote shares for each candidate, yielding 95\% CIs.

\textbf{Mean absolute error (MAE).} For $K$ candidates with
simulated shares $\hat{v}_k$ and actual shares $v_k$:
\begin{equation}
\text{MAE} = \frac{1}{K} \sum_{k=1}^{K} |\hat{v}_k - v_k|
\end{equation}

\textbf{Winner prediction.} Binary: 1 if
$\arg\max_k \hat{v}_k = \arg\max_k v_k$, else 0.

\textbf{Sido hit rate.} Fraction of \textit{sido} where the
predicted plurality winner matches the actual plurality winner.

\textbf{Wasserstein similarity.} We compute the 1-Wasserstein
distance between predicted and actual vote share distributions
across candidates, then convert to similarity:
\begin{equation}
W_{\text{sim}} = 1 - \frac{W_1(\hat{\mathbf{v}}, \mathbf{v})}{\max W_1}
\end{equation}
where $\max W_1$ is the maximum possible distance for the given
number of candidates. Values near 1 indicate close distributional
match.

\section{Experimental Setup}

\subsection{Elections and Population}

We evaluate Dynamo-K against six certified Korean elections
spanning eight years (Table~\ref{tab:elections}), with official
per-candidate and per-region vote shares obtained from the
National Election Commission's election statistics system
(선거통계시스템)~\cite{nec2025}. The set includes three
presidential elections, two general (National Assembly)
elections, and one local (광역단체장) election. The 2022 local
election (제8회 전국동시지방선거) is reserved as a true
cold held-out test for the OSLR calibration adapter
(Sec.~\ref{sec:oslr}): it is not used in any leave-one-election-out
cross-validation fold, and the adapter never sees its agents at
training time. Election margins span 0.73\%p to 17.1\%p.

\begin{table}[t]
\centering
\caption{Six backtested elections. Margins are between the top
  two candidates/parties. The 2022 local election is a cold
  held-out test for the OSLR adapter
  (Sec.~\ref{sec:oslr}).}
\label{tab:elections}
\small
\begin{tabular}{@{}lllr@{}}
\toprule
\textbf{Election} & \textbf{Date} & \textbf{Type} & \textbf{Margin} \\
\midrule
19th Presidential & 2017-05-09 & Pres. & 17.1\%p \\
21st General & 2020-04-15 & Gen. & 8.3\%p \\
20th Presidential & 2022-03-09 & Pres. & 0.73\%p \\
8th Local (광역단체장) & 2022-06-01 & Local & 10.0\%p \\
22nd General & 2024-04-10 & Gen. & 5.5\%p \\
21st Presidential & 2025-06-03 & Pres. & 8.3\%p \\
\bottomrule
\end{tabular}
\end{table}

All simulations use $N = 5{,}000$ agents with random seed 42 and
Gallup-calibrated orientations. The census baseline is fixed at
MDIS 2020 for all five elections; we acknowledge the temporal
mismatch for 2017 (3 years prior to census) and 2024--2025
(4--5 years after) as a limitation.

For presidential elections, agents vote for individual candidates.
For general elections, we simulate party-level vote preference
rather than constituency-level choices: agents vote for party
proxies (e.g., ``더불어민주당 후보'' / ``Democratic Party
candidate'') as a national-level abstraction. This design choice
reflects the infeasibility of modeling 254 individual constituency
races with 5{,}000 agents, but it loses constituency-specific
dynamics.

For the 2022 local election, each agent receives the named
광역단체장 (metropolitan mayor / provincial governor) candidate
slate for their sido (17 sidos, 54 candidates total across the
3 major parties plus minor-party entrants), and votes by name.
Sido-level results are aggregated to a national party-vote
share and a sido-hit-rate (parties winning at the sido level).
Because each agent's sido is already census-derived, no
constituency reassignment is needed.

\subsection{LLM Configuration}

All models are served locally via vLLM~\cite{vllm2023}. The
primary model Qwen3-30B-A3B is loaded with tensor parallelism
across two NVIDIA A100 80~GB GPUs. Key inference parameters:
temperature 0.5, maximum output tokens 300, JSON-mode
structured output.

For the cross-model robustness analysis (Sec.~\ref{sec:robustness}),
we additionally evaluate EXAONE-4.0-32B, a Korean-developed 32-billion
parameter model released by LG AI Research~\cite{exaone2025} with
substantial Korean-language pretraining coverage (its tokenizer
allocates Korean and English tokens in roughly equal proportions). We use
the official FP8-quantized checkpoint (\texttt{LGAI-EXAONE/EXAONE-4.0-32B-FP8})
served locally via vLLM on a single NVIDIA A100 80~GB GPU.
Llama-3.1-8B-Instruct and DeepSeek-R1-Distill-Llama-8B are also
served locally via vLLM on a single A100. Sampling parameters
match those used for Qwen3-30B-A3B.

The total GPU-hour cost per 5{,}000-agent simulation run is
approximately \$0.25 at typical on-demand A100 prices, against
industry-typical costs of order \$50{,}000 for a single
nationally-representative wave of professional CATI political
polling---two orders of magnitude lower marginal cost per
scenario explored. We note these
artifacts are not directly substitutable (Sec.~\ref{sec:cost}).

\section{Backtest Performance (Validation)}
\label{sec:prediction-validation}

We first report aggregate prediction performance across the six
backtested elections. We frame these numbers as validation evidence
that the simulation is calibrated well enough for the
mechanism-level diagnostics in Sec.~\ref{sec:mechanism} to be
informative---not as a polling replacement. Headline numbers
appear here; mechanism decomposition follows in
Sec.~\ref{sec:mechanism}; cost and limitations are deferred to
Sec.~\ref{sec:cost} and Sec.~\ref{sec:limitations}.

\subsection{Headline Backtest Numbers}
\label{sec:overall-accuracy}

Table~\ref{tab:results} presents the full backtest results across
all five elections used for the leave-one-election-out CV pool;
the held-out 2022 local election is reported separately in
Sec.~\ref{sec:local-holdout}.

\begin{table}[t]
\centering
\caption{Backtest results across five Korean elections. MAE is
  mean absolute error in percentage points. Winner indicates
  correct (\checkmark) or incorrect ($\times$) prediction of the
  plurality winner. Sido hit is the fraction of provinces where
  the predicted winner matches actual. $W_{\text{sim}}$ is
  Wasserstein similarity. Abstention is simulated abstention rate.}
\label{tab:results}
\small
\begin{tabular}{@{}llccccc@{}}
\toprule
\textbf{Election} & \textbf{Type} & \textbf{MAE (\%p)} &
  \textbf{Winner} & \textbf{Sido Hit} & $\boldsymbol{W_{\textbf{sim}}}$ &
  \textbf{Abstain} \\
\midrule
19th Presidential 2017 & Pres. & 13.3 & \checkmark & 82.4\% & 0.87 & 0.8\% \\
21st General 2020 & Gen. & 5.3 & $\times$ & 64.7\% & 0.95 & 7.0\% \\
20th Presidential 2022 & Pres. & 2.1 & \checkmark & 82.4\% & 0.98 & 0.6\% \\
22nd General 2024 & Gen. & 4.4 & $\times$ & 60.0\% & 0.99 & 6.4\% \\
21st Presidential 2025 & Pres. & 7.1 & \checkmark & 73.3\% & 0.93 & 1.1\% \\
\midrule
\textbf{Average} & & \textbf{6.5} & \textbf{60\%} & \textbf{72.6\%} &
  \textbf{0.94} & \textbf{3.2\%} \\
\quad Presidential avg. & & 7.5 & 3/3 & 79.4\% & 0.93 & 0.8\% \\
\quad General avg. & & 4.9 & 0\% & 62.4\% & 0.97 & 6.7\% \\
\bottomrule
\end{tabular}
\end{table}

The headline finding is a stark split between election types:
presidential winner prediction is 3/3 correct (binomial 95\% CI
[29.2\%, 100\%]), while
general election winner prediction is 0\% (2/2 incorrect). The
overall average MAE is 6.5\%p, with presidential elections
averaging 7.5\%p (inflated by the 2017 outlier) and general
elections averaging 4.9\%p. Wasserstein similarity is consistently
high (0.87--0.99), indicating that predicted vote distributions
track actual distributions well even when the winner prediction
fails.

\subsection{Presidential Election Performance}

Table~\ref{tab:candidates} provides a detailed per-candidate
comparison of simulated and actual vote shares with bootstrap 95\%
confidence intervals for all three presidential elections.

\begin{table}[t]
\centering
\caption{Per-candidate vote shares (\%) for presidential elections.
  Sim = simulated share, 95\% CI = bootstrap confidence interval
  (1{,}000 resamples), Act = actual certified share,
  Err = signed error (Sim $-$ Act).}
\label{tab:candidates}
\small
\begin{tabular}{@{}llrcrr@{}}
\toprule
\textbf{Election} & \textbf{Candidate (Party)} &
  \textbf{Sim} & \textbf{95\% CI} & \textbf{Act} & \textbf{Err} \\
\midrule
\multirow{5}{*}{19th Pres.\ 2017}
  & 문재인 / Moon (더불어민주당) & 71.3 & [70.0, 72.5] & 41.1 & +30.2 \\
  & 홍준표 / Hong (자유한국당) & 27.4 & [26.2, 28.7] & 24.0 & +3.4 \\
  & 안철수 / Ahn (국민의당) & 0.9 & [0.6, 1.1] & 21.4 & $-$20.5 \\
  & 유승민 / Yoo (바른정당) & 0.5 & [0.3, 0.7] & 6.8 & $-$6.3 \\
  & 심상정 / Sim (정의당) & 0.02 & [0.0, 0.06] & 6.2 & $-$6.2 \\
\midrule
\multirow{3}{*}{20th Pres.\ 2022}
  & 윤석열 / Yoon (국민의힘) & 52.3 & [50.9, 53.7] & 48.6 & +3.7 \\
  & 이재명 / Lee (더불어민주당) & 47.6 & [46.2, 49.0] & 47.8 & $-$0.2 \\
  & 심상정 / Sim (정의당) & 0.06 & [0.0, 0.12] & 2.4 & $-$2.3 \\
\midrule
\multirow{2}{*}{21st Pres.\ 2025}
  & 이재명 / Lee (더불어민주당) & 61.3 & [59.9, 62.7] & 49.4 & +11.8 \\
  & 김문수 / Kim (국민의힘) & 38.7 & [37.3, 40.1] & 41.2 & $-$2.4 \\
\bottomrule
\end{tabular}
\end{table}

\textbf{20th Presidential 2022: Best case.} This election produced
our strongest result: 2.1\%p MAE on a race decided by just
0.73\%p. The simulated shares were 윤석열 (Yoon Suk Yeol)
52.3\% vs.\ 이재명 (Lee Jae-myung) 47.6\%, compared to actual
results of 48.6\% vs.\ 47.8\% (Table~\ref{tab:candidates}). The
system correctly predicted the conservative winner despite the
extremely tight margin. The largest per-candidate error was 3.7\%p
for Yoon, indicating a slight overestimation of conservative
support that was nonetheless insufficient to flip the predicted
winner. The \textit{sido} hit rate of 82.4\% (14/17 provinces)
indicates strong sub-national calibration, with errors concentrated
in swing regions. Notably, 심상정 (Sim Sang-jeong) of the Justice
Party was simulated at 0.06\% vs.\ her actual 2.4\%, continuing
the minor-candidate compression pattern. The simulated abstention
rate was just 0.6\%, consistent with the high-salience nature of
Korean presidential contests.

\textbf{21st Presidential 2025: Crisis election.} The 2025 snap
election, triggered by President Yoon Suk Yeol's impeachment and
removal from office, tested the system on an unprecedented scenario
with no historical parallel in Korean democracy. The simulation
predicted 이재명 (Lee Jae-myung) 61.3\% vs.\ 김문수 (Kim Moon-soo)
38.7\%, against actual results of 49.4\% vs.\ 41.2\%, yielding
7.1\%p MAE. The winner was correctly identified, but the margin was
substantially overpredicted: simulated margin of 22.6\%p vs.\ actual
8.3\%p. This overprediction is partly explained by the system not
modeling the third candidate 이준석 (Lee Jun-seok) of the Reform
Party, who received 8.3\% of actual votes. His voters likely came
disproportionately from the moderate-to-conservative pool, and
their absence from the simulation inflated Lee Jae-myung's share.
The \textit{sido} hit rate was 73.3\% (11/15 reporting regions),
lower than the 2022 rate, reflecting the more volatile political
environment.

\textbf{19th Presidential 2017: Worst case.} This five-candidate
race produced the highest MAE (13.3\%p) due to severe third-party
candidate collapse. The simulation predicted 문재인 (Moon Jae-in)
at 71.3\%, far above his actual 41.1\% (+30.2\%p error), while
안철수 (Ahn Cheol-soo) was simulated at just 0.9\% against an
actual 21.4\% ($-$20.5\%p error). The remaining third-party
candidates 유승민 (Yoo Seung-min, 6.8\% actual) and 심상정 (Sim
Sang-jeong, 6.2\% actual) were similarly compressed to near-zero.
The system effectively collapsed a competitive multi-candidate
field into a two-horse race, concentrating votes on the eventual
winner at the expense of third-party candidates. Despite this
distortion, the winner was correctly predicted, and the \textit{sido}
hit rate remained high at 82.4\%, suggesting that the geographic
pattern of support was accurate even as the magnitude was not.

\begin{figure*}[t]
\centering
\begin{tikzpicture}
\begin{axis}[
    ybar,
    width=\textwidth,
    height=6.5cm,
    bar width=6pt,
    ylabel={Vote share (\%)},
    symbolic x coords={
      {Moon 2017},{Hong 2017},{Ahn 2017},
      {Yoon 2022},{Lee 2022},
      {Lee 2025},{Kim 2025}},
    xtick=data,
    x tick label style={rotate=35, anchor=east, font=\small},
    ymin=0, ymax=80,
    legend style={at={(0.02,0.98)}, anchor=north west, font=\small},
    enlarge x limits=0.1,
    nodes near coords style={font=\tiny, rotate=90, anchor=west},
    every node near coords/.append style={xshift=0pt},
    ymajorgrids=true,
    grid style=dashed,
    title style={font=\normalsize},
]
\addplot[fill=blue!60, nodes near coords]
  coordinates {
    ({Moon 2017},71.3) ({Hong 2017},27.4) ({Ahn 2017},0.9)
    ({Yoon 2022},52.3) ({Lee 2022},47.6)
    ({Lee 2025},61.3) ({Kim 2025},38.7)
  };
\addplot[fill=red!50, nodes near coords]
  coordinates {
    ({Moon 2017},41.1) ({Hong 2017},24.0) ({Ahn 2017},21.4)
    ({Yoon 2022},48.6) ({Lee 2022},47.8)
    ({Lee 2025},49.4) ({Kim 2025},41.2)
  };
\legend{Simulated, Actual}
\end{axis}
\end{tikzpicture}
\caption{Predicted vs.\ actual vote shares for the three
  presidential elections. The 2022 election shows the closest
  match (2.1\%p MAE). The 2017 election shows severe third-party
  collapse: 안철수 (Ahn) received 21.4\% actual but only 0.9\%
  simulated, with votes redistributed primarily to 문재인 (Moon).}
\label{fig:barplot}
\end{figure*}
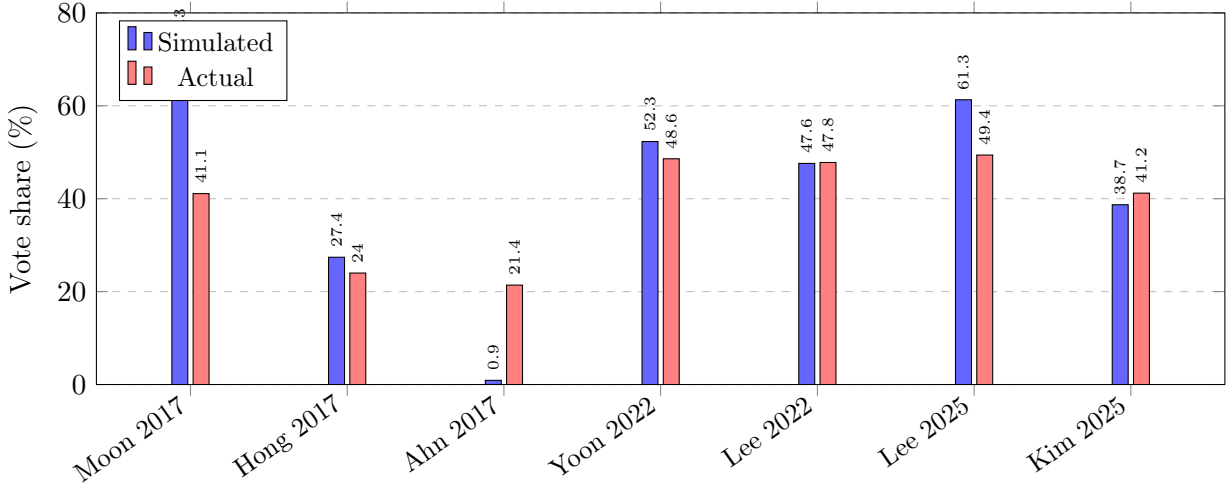

Figure~\ref{fig:barplot} visualizes the predicted vs.\ actual
vote shares for all three presidential elections. The 2022 result
is striking in its proximity to actual outcomes; the 2017 result
visually demonstrates the third-party collapse phenomenon.

\subsection{General Election Performance}

Both general election simulations predicted conservative party
victories, while the actual winner in both cases was the
progressive Democratic Party (더불어민주당). In the 2020 election,
simulated shares were 미래통합당 (conservative) 46.4\% vs.\
더불어민주당 (progressive) 45.5\%, against actual results of
41.5\% vs.\ 49.8\% (MAE 5.3\%p). In 2024, simulated shares were
국민의힘 (conservative) 50.0\% vs.\ 더불어민주당 47.0\%, against
actual results of 45.0\% vs.\ 50.5\% (MAE 4.4\%p).

Two structural features distinguish general from presidential
simulations. First, simulated abstention rates are substantially
higher for general elections (6.4--7.0\%) than for presidential
elections (0.6--1.1\%), reflecting agents' difficulty engaging with
party-proxy candidates lacking individual identities. Second, the
party-proxy abstraction loses constituency-specific dynamics:
local candidate quality, incumbency advantages, and tactical
voting patterns that collectively determine general election
outcomes at the district level.

Both general elections exhibit a consistent conservative bias of
approximately 3--5\%p in the simulation, suggesting a systematic
calibration issue specific to the party-proxy framing rather than
random error.

\subsection{Held-Out 2022 Local Election}
\label{sec:local-holdout}

The 2022 local election (제8회 전국동시지방선거, 광역단체장) is
reserved as a cold held-out test set: it is not used in any
leave-one-election-out fold, and the OSLR adapter
(Sec.~\ref{sec:oslr}) is trained without ever seeing 2022-local
agents. Structurally this election differs from both prior
backtests---it is named-candidate (like presidential) but
parallel-race (like general), with 17 광역단체장 contests, each
sido carrying its own slate of 2--5 candidates and three majors
in every race (Table~\ref{tab:local-2022}).

The actual outcome was a People Power Party landslide:
국민의힘 won 12 of 17 sido (서울, 부산, 대구, 인천, 대전,
울산, 강원, 충남, 충북, 경남, 경북, 제주) and 더불어민주당
won 5 (광주, 세종, 경기, 전남, 전북), with national party
vote share 54.0\% vs.\ 44.0\%---a 10.0\%p margin and the
strongest sustained signal of post-2022 anti-Democratic
sentiment until the eventual Yoon impeachment.

\begin{table}[t]
\centering
\caption{Held-out 2022 local election (광역단체장): simulated
  vs.\ actual party vote share and sido-hit rate. ``Sidos won''
  reports the count of the 17 sido-level races in which the model
  predicts each major party as winner; ``Nat.\ MAE'' is the mean
  absolute error of seat-share against the actual; ``Sido hit'' is
  the count of sidos where the predicted winning party matches
  the actual.}
\label{tab:local-2022}
\small
\begin{tabular}{@{}lccrr@{}}
\toprule
Model & Sidos won (DPK/PPP) & Nat.\ MAE & Sido hit & Pred winner \\
\midrule
Qwen3-30B-A3B   & 12/5 & 41.2\%p & 10/17 & 더불어민주당 \\
Llama-3.1-8B    & 14/3 & 52.9\%p & 8/17  & 더불어민주당 \\
EXAONE-4.0-32B  & 5/12 & 0.0\%p  & 15/17 & 국민의힘 \\
DeepSeek-R1-8B  & 7/10 & 11.8\%p & 13/17 & 국민의힘 \\
\midrule
\textbf{Actual} & \textbf{5/12} & \textbf{0} & \textbf{17/17} & \textbf{국민의힘} \\
\bottomrule
\end{tabular}
\end{table}

Two structural features make this a useful held-out probe.
First, named-candidate prompting eliminates the party-proxy
abstraction that drove the 0/2 general-election failure
(Sec.~\ref{sec:gen-bias}): each sido's prompt uses the actual
candidates (e.g., 송영길 vs.\ 오세훈 vs.\ 권수정 for 서울).
Second, the OSLR adapter's $\beta_{o,s}$ (orientation $\times$
sido) parameters are \emph{defined} on sido-level variation, so
a single-race-per-sido design is the cleanest measurement of
that signal without presidential coattail or sungeo dilution.

\subsection{Sub-National Accuracy}
\label{sec:sub-national-accuracy}

Sido hit rates range from 60.0\% to 82.4\%, with presidential
elections consistently outperforming general elections (79.4\%
vs.\ 62.4\% average). The simulation correctly reproduces the
fundamental geographic pattern of Korean politics: 경상
(Gyeongsang) provinces vote conservative and 전라 (Jeolla)
provinces vote progressive in all five simulations.

Errors concentrate in swing regions---서울 (Seoul), 경기
(Gyeonggi), and 충청 (Chungcheong)---where moderate voters
dominate and small shifts in their behavior determine the
local winner. This is consistent with the broader pattern:
the system performs best where political orientation is
strongly determined by demographics (regional strongholds)
and worst where it depends on campaign-specific factors
(swing regions).

Table~\ref{tab:regional} compares simulated and KGSS-derived
orientation distributions for selected regions, confirming that
the census grounding and KGSS conditioning successfully
reproduce Korea's regional political geography.

\begin{table}[t]
\centering
\caption{Simulated vs.\ KGSS orientation distribution (\%)
  for selected regions. Con = conservative, Mod = moderate,
  Pro = progressive.}
\label{tab:regional}
\small
\begin{tabular}{@{}lcccccc@{}}
\toprule
& \multicolumn{3}{c}{\textbf{Simulated}} &
  \multicolumn{3}{c}{\textbf{KGSS}} \\
\cmidrule(lr){2-4} \cmidrule(lr){5-7}
\textbf{Region} & Con & Mod & Pro & Con & Mod & Pro \\
\midrule
경상 (Gyeongsang) & 34 & 44 & 22 & 38 & 32 & 30 \\
전라 (Jeolla) & 14 & 50 & 36 & 16 & 28 & 56 \\
수도권 (Capital) & 24 & 50 & 26 & 28 & 34 & 38 \\
충청 (Chungcheong) & 28 & 48 & 24 & 30 & 34 & 36 \\
\bottomrule
\end{tabular}
\end{table}

\subsection{Aggregate Subgroup Validation vs.\ Exit Polls}
\label{sec:subgroup}

Beyond national and \textit{sido}-level accuracy, we validate
simulation outputs against the KBS/MBC/SBS joint broadcast exit
polls for the three presidential elections. These polls, with
$N\approx 80{,}000$ and quoted precision of $\pm 0.8$\%p, report
vote shares broken down by age~$\times$~sex cells and serve as
an independent aggregate benchmark---the strongest external
subgroup check available without access to individual-level
survey microdata. Table~\ref{tab:subgroup-validation} reports
cell-level MAE and Pearson correlation between simulated
cell shares and exit-poll cell shares.

\begin{table}[t]
\centering
\caption{Aggregate subgroup validation: simulated age$\times$sex
  (or age-only for 2017) vote shares versus joint broadcast exit
  polls. Cell MAE is averaged across (cell, candidate) pairs;
  $r$ is Pearson correlation across cells.}
\label{tab:subgroup-validation}
\small
\begin{tabular}{@{}llccr@{}}
\toprule
Election & Model & Cell MAE & $r$ & $N$ \\
\midrule
19th Pres.\ 2017 & Qwen3-30B       & 14.7 & 0.82 & 25 \\
19th Pres.\ 2017 & Llama-3.1-8B    & 14.9 & 0.83 & 25 \\
19th Pres.\ 2017 & EXAONE-4.0-32B  & 13.9 & 0.83 & 25 \\
19th Pres.\ 2017 & DeepSeek-R1-8B  & 9.6  & 0.77 & 25 \\
\midrule
20th Pres.\ 2022 & Qwen3-30B       & 7.8  & 0.92 & 36 \\
20th Pres.\ 2022 & Llama-3.1-8B    & 15.7 & 0.70 & 36 \\
20th Pres.\ 2022 & EXAONE-4.0-32B  & 6.4  & 0.93 & 36 \\
20th Pres.\ 2022 & DeepSeek-R1-8B  & 22.9 & 0.53 & 36 \\
\midrule
21st Pres.\ 2025 & Llama-3.1-8B    & 13.0 & 0.70 & 36 \\
21st Pres.\ 2025 & EXAONE-4.0-32B  & 13.4 & 0.68 & 36 \\
21st Pres.\ 2025 & DeepSeek-R1-8B  & 16.4 & 0.48 & 36 \\
\bottomrule
\end{tabular}
\end{table}

Two findings stand out. First, Pearson correlations reach
$r=0.93$ for EXAONE and $r=0.92$ for Qwen3 on the 2022 race,
indicating that the census-grounded pipeline captures the
\emph{direction} of subgroup variation even where absolute
cell MAE is high. Second, the celebrated MZ gender gap
(이대남/이대녀) is reproduced directionally but substantially
muted in magnitude: the 2022 exit poll reports a 24.9\%p
gap between 20대 남성 and 20대 여성 윤석열 vote share, while
the best-calibrated simulations (Qwen3 and EXAONE) produce
only 6.0--6.8\%p gaps. A similar attenuation appears in 2025,
where 이준석 received 37.2\% from 20대 남성 vs.\ 10.3\% from
20대 여성---a gap the simulations capture in direction but not
magnitude. The 2025 KBS/MBC/SBS exit poll itself overestimated
이재명 by roughly 2.3\%p nationally (social-desirability pressure
on 김문수 voters), so we do not treat it as ground truth but
as an independent benchmark of known precision.

\subsection{Where the General-Election Bias Comes From}
\label{sec:gen-bias}

The 0/2 general-election winner accuracy in
Section~\ref{sec:sub-national-accuracy} is sometimes described
as a ``conservative bias.'' Decomposing the error across
models reveals a more precise diagnosis: the dominant failure
is a \textbf{regional polarization collapse} rather than a
uniform conservative tilt. Averaged across the four models,
simulation systematically \emph{underpredicts} the actual
winner in both 전라 strongholds (전라남도 $-29$--$-32$\%p
vs.\ the Democratic winner in 2020/2024) and 경상/대구
strongholds (대구광역시 $-25$--$-28$\%p vs.\ the conservative
winner), while modestly overpredicting winners in swing
regions. The national tilt appears conservative only because
경상 contains a larger electorate than 전라.

Turnout reweighting provides a second diagnostic lever.
Using NEC-published age-bracket turnout rates, we rescale each
agent's contribution by their bracket's turnout
propensity---aligning the simulation's uniform near-99\%
participation with the actual 66--67\% turnout skewed toward
older (and therefore more conservative) voters.
Table~\ref{tab:general-bias-decomp} reports raw vs.\
turnout-weighted MAE for each (election, model) pair.
Reweighting reduces MAE on six of eight runs, most strongly
for EXAONE-4.0-32B ($-0.64$\%p on 2020, $-0.76$\%p on 2024) and
DeepSeek-R1-8B ($-0.32$ to $-0.38$\%p). The effect is small but
consistent, suggesting that a meaningful fraction of the
residual general-election error is composition-driven and
that adaptive turnout modeling is a cheap, principled remedy.

\begin{table}[t]
\centering
\caption{General-election MAE decomposition: raw vs.\ turnout
  reweighted. Negative $\Delta$ = reweighting helped.}
\label{tab:general-bias-decomp}
\small
\begin{tabular}{@{}llccccc@{}}
\toprule
Election & Model & Raw & R.W & $\Delta$ & R.Win & TW.Win \\
\midrule
21st Gen.\ 2020 & Qwen3-30B      & 5.3  & 5.7  & $+0.4$ & $\times$ & $\times$ \\
21st Gen.\ 2020 & Llama-3.1-8B   & 10.8 & 10.4 & $-0.4$ & \checkmark & \checkmark \\
21st Gen.\ 2020 & EXAONE-4.0-32B & 4.1  & 3.5  & $-0.6$ & \checkmark & \checkmark \\
21st Gen.\ 2020 & DeepSeek-R1-8B & 5.0  & 4.7  & $-0.4$ & \checkmark & \checkmark \\
22nd Gen.\ 2024 & Qwen3-30B      & 4.4  & 5.0  & $+0.6$ & $\times$ & $\times$ \\
22nd Gen.\ 2024 & Llama-3.1-8B   & 5.8  & 6.3  & $+0.5$ & $\times$ & $\times$ \\
22nd Gen.\ 2024 & EXAONE-4.0-32B & 3.3  & 2.6  & $-0.8$ & \checkmark & \checkmark \\
22nd Gen.\ 2024 & DeepSeek-R1-8B & 1.8  & 1.5  & $-0.3$ & \checkmark & \checkmark \\
\bottomrule
\end{tabular}
\end{table}

\section{Prediction Performance as Validation, Cost, and Limitations}
\label{sec:discussion}
\label{sec:mechanism}

We now report prediction performance and its interpretation.
We frame these results as validation evidence that the diagnostic
apparatus of Sec.~\ref{sec:mechanism} operates on a simulation
that is well-enough calibrated to produce informative
mechanism-level signal---not as a polling replacement. The cost
analysis (Sec.~\ref{sec:cost}) is correspondingly reframed as
``diagnosis at scale'' rather than ``cheap polling.''

\subsection{LLM Progressive Bias and Mitigation}
\label{sec:bias-ablation}

Our experience with LLM progressive bias is thematically consistent
with Feng et al.'s~\cite{columbia2025} finding of a liberal lean in
GPT-family LLMs, though their methodology differs (intrinsic
Political Compass probing vs.\ our voter-decision simulation). In
our initial configuration, moderate agents---48\% of the simulated
electorate---voted for progressive candidates 97\% of the time.
This bias was catastrophic for prediction: it made conservative
victories structurally impossible regardless of the actual
political environment.

Our mitigation strategy combined four prompt engineering
techniques: (1)~explicit party alignment cues naming specific
Korean parties in orientation descriptions, (2)~balanced moderate
voter descriptions emphasizing equal voting probability for both
camps, (3)~the ``AI의 관점이 아닌'' (``not from the AI's
perspective'') instruction targeting the model's tendency to
express its own values, and (4)~reinforced orientation-consistent
voting instructions in the user prompt.

\textbf{Ablation against a vanilla prompt.}
To quantify the mitigation block's contribution, we re-ran all
five backtest elections on Qwen3-30B-A3B and EXAONE-4.0-32B with
a stripped \emph{vanilla} prompt---a single ``pretend to be a
Korean voter, your demographics are X, choose a candidate''
instruction---and compared the result to the full Dynamo-K
pipeline (Table~\ref{tab:vanilla-ablation}). Across the ten
(model, election) cells, the vanilla prompt averages
\textbf{36.8\%p MAE} versus \textbf{7.1\%p} for the full
pipeline, a \textbf{5.2$\times$} reduction. The orientation--vote
consistency measure (``Mod$\to$Prog'') tracks this: moderate
agents vote for the 더불어민주당-lineage progressive candidate
\textbf{82\%} of the time under the vanilla prompt versus
\textbf{59\%} under the full pipeline, much closer to the
swing-voter behavior implied by Gallup data. Notably, vanilla
winner accuracy (7/10) is identical to full-pipeline winner
accuracy---vanilla is ``right for the wrong reason'': it
predicts large progressive landslides that happen to coincide
with progressive victories, but with vote-share errors so large
that close races (2022, 2025) become uncallable.

\begin{table}[t]
\centering
\caption{Ablation: full Dynamo-K pipeline vs.\ vanilla
``pretend to be a Korean voter'' prompt across two models and
five elections. ``Mod$\to$Prog'' is the fraction of
moderate-orientation agents voting for the progressive candidate
(the 더불어민주당 lineage); the vanilla rate of 82\% averaged
over ten cells is the pre-mitigation behavior the bias block was
designed to fix.}
\label{tab:vanilla-ablation}
\small
\begin{tabular}{llcccccc}
\toprule
 & & \multicolumn{2}{c}{MAE (\%p)} & \multicolumn{2}{c}{Winner} & \multicolumn{2}{c}{Mod$\to$Prog} \\
\cmidrule(lr){3-4} \cmidrule(lr){5-6} \cmidrule(lr){7-8}
Election & Model & Full & Van.\ & Full & Van.\ & Full & Van.\ \\
\midrule
19 Pres.\ 2017 & Qwen3   & 13.3 & 31.6 & \checkmark & \checkmark & 95\% & 99\% \\
               & EXAONE  & 12.5 & 18.8 & \checkmark & \checkmark & 91\% & 75\% \\
\addlinespace
21 Gen.\ 2020  & Qwen3   & 5.3  & 46.5 & $\times$   & \checkmark & 56\% & 100\% \\
               & EXAONE  & 4.1  & 44.8 & \checkmark & \checkmark & 67\% & 99\% \\
\addlinespace
20 Pres.\ 2022 & Qwen3   & 2.1  & 48.2 & \checkmark & $\times$   & 45\% & 98\% \\
               & EXAONE  & 6.4  & 26.1 & \checkmark & $\times$   & 28\% & 86\% \\
\addlinespace
22 Gen.\ 2024  & Qwen3   & 4.4  & 49.1 & $\times$   & $\times$   & 44\% & 2\% \\
               & EXAONE  & 3.3  & 46.4 & \checkmark & \checkmark & 63\% & 100\% \\
\addlinespace
21 Pres.\ 2025 & Qwen3   & 7.1  & 15.5 & \checkmark & \checkmark & 74\% & 70\% \\
               & EXAONE  & 12.5 & 41.3 & $\times$   & \checkmark & 32\% & 96\% \\
\midrule
\textbf{Avg} & both & \textbf{7.1} & \textbf{36.8} & \textbf{7/10} & \textbf{7/10} & \textbf{59\%} & \textbf{82\%} \\
\bottomrule
\end{tabular}
\end{table}

This reduction demonstrates that prompt engineering can
substantially mitigate LLM political bias but does not fully
resolve it. The 59\% moderate-progressive rate, while dramatically
improved over both the 97\% pre-mitigation baseline and the 82\%
vanilla rate, still depends on prompt-specific calibration. We
recommend that any LLM electoral simulation system include
systematic measurement of orientation--vote consistency against
external benchmarks as a standard validation step.

An important open question is whether the bias is
language-dependent. Our Korean-language prompts may elicit
different bias patterns than English-language prompts, as the
political content of Korean-language training data has a
different ideological distribution than English-language
political text. A systematic cross-lingual comparison is needed
to understand this dimension.

\subsection{Third-Party Candidate Collapse}
\label{sec:third-party}

The 2017 presidential election reveals a fundamental failure
mode: LLM agents struggle to distribute votes among multiple
viable candidates. In a five-candidate field, the simulation
concentrated 98.6\% of votes on the top two (문재인 71.3\%,
홍준표 27.4\%), while the actual top-two share was only 65.1\%
(문재인 41.1\%, 홍준표 24.0\%). The centrist 안철수 received
21.4\% of actual votes but only 0.9\% in simulation.

We hypothesize two contributing mechanisms. First,
\textit{hindsight bias in world knowledge}: LLMs are trained on
text written after elections, including narratives that portray
the eventual winner as the dominant figure and treat third-party
candidates as marginal. This post-hoc framing likely influences
the model's prior probability assignments. Second,
\textit{ideological simplification}: the system's three-way
orientation (progressive/moderate/conservative) naturally maps
to a two-party contest, lacking the resolution to capture voters
whose primary allegiance is to a centrist ``third way'' rather
than to either major camp.

A similar but milder pattern appears in the 2025 election, where
이준석 (Lee Jun-seok) received 8.3\% of actual votes but was
not included in the simulation. Modeling third-party dynamics
likely requires either explicitly including all significant
candidates in the ballot (with appropriate name recognition
calibration) or developing a more nuanced orientation space that
can represent centrist and protest-vote motivations. As we show
below (Table~\ref{tab:reframed-2017}), the salience of the
third candidate within the scenario prompt itself is the
dominant lever; pure ``hindsight bias in world knowledge'' is
a weaker contributor than the two-mechanism framing above
implies.

\subsubsection{Mechanism: salience failure vs.\ decision bias}
\label{sec:third-party-mechanism}

To distinguish whether the collapse is driven by (a) the LLM
forgetting the third candidate exists (salience failure) or
(b) the LLM remembering the candidate but not choosing them
(decision bias), we regex-classify every agent's reasoning
text for candidate mentions across all four models on the
2017 election.
Table~\ref{tab:third-party-mechanism} reports, for each of
안철수/유승민/심상정, the actual vote share, the simulated
share, the mention rate in reasoning text, and the
conditional vote rate given mention.

\begin{table}[t]
\centering
\caption{2017 third-party collapse mechanism. Sim = simulated
  vote share. Ment = fraction of non-abstaining agents whose
  reasoning text contains the candidate's name. V$\mid$M =
  among those who mention the candidate, the fraction who
  actually voted for them. High Ment with low V$\mid$M = decision
  bias; low Ment = salience failure.}
\label{tab:third-party-mechanism}
\small
\setlength{\tabcolsep}{3pt}
\begin{tabular}{@{}lrrrr@{}}
\toprule
Cand. (actual \%) & Model & Sim & Ment & V$\mid$M \\
\midrule
\multirow{4}{*}{안철수 (21.4)}
 & Qwen3-30B       & 0.9 & 10.6 & 7.4 \\
 & Llama-3.1-8B    & 1.7 & 4.8  & 34.4 \\
 & EXAONE-4.0-32B  & 0.9 & 0.7  & 97.1 \\
 & DeepSeek-R1-8B  & 9.2 & 22.0 & 34.5 \\
\midrule
\multirow{4}{*}{유승민 (6.8)}
 & Qwen3-30B       & 0.5 & 2.8  & 16.3 \\
 & Llama-3.1-8B    & 0.1 & 1.7  & 5.8 \\
 & EXAONE-4.0-32B  & 2.0 & 1.4  & 100.0 \\
 & DeepSeek-R1-8B  & 5.1 & 23.0 & 21.1 \\
\midrule
\multirow{4}{*}{심상정 (6.2)}
 & Qwen3-30B       & 0.0 & 6.9  & 0.3 \\
 & Llama-3.1-8B    & 0.0 & 1.6  & 2.5 \\
 & EXAONE-4.0-32B  & 0.4 & 0.4  & 94.7 \\
 & DeepSeek-R1-8B  & 0.7 & 10.3 & 5.2 \\
\bottomrule
\end{tabular}
\end{table}

The mechanism is \emph{heterogeneous across models}. EXAONE
and (for 유승민/심상정) Llama exhibit near-total salience
failure---fewer than 2\% of agents mention these candidates
at all---but when they do, they overwhelmingly vote for them
(94--100\% for EXAONE). Qwen3 and DeepSeek-R1 show the opposite
pattern: both models \emph{repeatedly} reason about the
third-party candidates (Qwen3 mentions 안철수 in 10.6\% of
agent reasonings, DeepSeek in 22.0\%) but funnel that
consideration back to the Big Two majors at the decision step,
achieving V$\mid$M rates of 7--35\%.
DeepSeek-R1-8B, the only reasoning-chain-trained model in the
set, produces the least-collapsed prediction (9.2\% for
안철수), consistent with the hypothesis that explicit
deliberation partially counteracts the LLM hindsight pull
toward the eventual winner.

\paragraph{Scenario reframing recovers most of the collapsed share.}
To test whether the collapse is driven by knowledge (the LLM
``knows'' 안철수 lost and biases against him) or by framing (the
scenario prompt fails to make him salient), we rebuilt the 2017
\texttt{ElectionScenario} as a \emph{reframed} variant on
Qwen3-30B-A3B: candidates are reordered to put 안철수 first, his
pledges are expanded to match the level of detail given to the
two major candidates, and the context paragraph emphasizes the
three-way 문/안/홍 narrative present in pre-election polling.
Crucially, no hidden labels or vote nudges are added---only
information that was already public before election day.
Table~\ref{tab:reframed-2017} shows the effect: the 안철수
predicted share moves from 0.9\% to \textbf{18.8\%} against an
actual 21.4\%, and election-level MAE drops from
\textbf{13.3\%p to 5.1\%p} (a 62\% reduction). The recovery comes
almost entirely out of 문재인's overshoot, which falls from
71.2\% (default) to 52.7\% (reframed; actual 41.1\%).

\begin{table}[t]
\centering
\caption{Default vs.\ reframed 2017 scenario on Qwen3-30B-A3B.
Reframed reorders the candidate list to put 안철수 first, expands
his pledges, and rewrites the context paragraph to emphasize the
three-way 문/안/홍 narrative. Recovery of the 안철수 share from
$\sim 1\%$ to ${\sim}19\%$ supports a framing-driven (rather than
knowledge-driven) interpretation of the 2017 third-party collapse.}
\label{tab:reframed-2017}
\small
\begin{tabular}{lcccc}
\toprule
Candidate & Actual & Default & Reframed & $\Delta$ (R$-$D) \\
\midrule
문재인 & 41.1\% & 71.2\% & 52.7\% & $-$18.6 \\
홍준표 & 24.0\% & 27.4\% & 26.3\% & $-$1.1 \\
안철수 & 21.4\% &  0.9\% & 18.8\% & $+$17.9 \\
유승민 &  6.8\% &  0.5\% &  0.0\% & $-$0.5 \\
심상정 &  6.2\% &  0.0\% &  2.3\% & $+$2.2 \\
\midrule
\textbf{MAE (\%p)} & --- & \textbf{13.3} & \textbf{5.1} & $\mathbf{-8.2}$ \\
\bottomrule
\end{tabular}
\end{table}

This is informative because Qwen3 itself shows the
\emph{decision-bias} pattern in
Table~\ref{tab:third-party-mechanism} (10.6\% mention rate but
only 7.4\% V$\mid$M)---i.e., the default scenario was not failing
because Qwen3 lacked 안철수 from its world knowledge, but because
the prompt's framing did not legitimize him as a viable choice.
Reframing alone substantially closes the gap without any
reasoning-chain architecture, revising our earlier conjecture:
explicit deliberation (DeepSeek-R1) helps, but scenario framing
is the larger lever. The salience-failure cluster (EXAONE; Llama
on 유승민/심상정) is a separate failure mode that reframing
cannot fix---those models would additionally need the third
candidate's biographical content surfaced into the agent prompt.

\subsection{Presidential vs.\ General Election Accuracy}

The starkest pattern in our results is the divergence between
presidential (3/3 correct) and general (0/2 correct) elections. Paradoxically, the general election MAE
(4.9\%p average) is lower than the presidential MAE (7.5\%p
average), indicating that the vote share predictions are closer
but systematically biased in the wrong direction.

We attribute this to the party-proxy abstraction used for general
elections. Presidential elections are personal contests: voters
evaluate specific candidates, and the simulation's agent personas
can meaningfully engage with individual candidate attributes.
General elections, by contrast, are 254 parallel constituency
contests where local candidate quality, incumbency, and tactical
voting matter enormously. Our party-proxy approach
(``더불어민주당 후보'' / ``Democratic Party candidate'' without
individual identity) strips these dynamics, leaving only
ideological preference---which, as we have shown, exhibits a
residual conservative bias of 3--5\%p.

The practical prescription is clear: general election simulation
requires constituency-level modeling with specific candidates,
local issue contexts, and potentially different agent population
sizes per district. This is substantially more complex than
presidential simulation---scaling from 1 national contest to 254
constituency races---and represents the primary direction for
future work.

\paragraph{Constituency-level proof of concept (2024 with Qwen3).}
The constituency variant was developed to test whether the
party-proxy abstraction or the underlying simulation was the
bottleneck behind the 0/2 general-election failure; we report it
here as a methodological proof of concept rather than as a
retrofit of the headline backtest in Table~\ref{tab:results}.
We implement a constituency-level variant of the pipeline for
the 22nd General Election (2024) using Qwen3-30B-A3B: for each
of the 244 single-member districts we build an
\texttt{ElectionScenario} from NEC-registered candidates for
that district, allocate agents proportionally by elector count,
and aggregate per-district winners into national seat counts
(Table~\ref{tab:constituency-backtest}). The constituency-level
run attains 67.6\% district accuracy (165 of 244 correct), with
100\% accuracy in partisan-stronghold sidos (경상북도, 광주광역시,
세종, 제주) and degraded performance in swing sidos
(충청남도 18\%, 강원/충북 25\%). Crucially, aggregating
predicted district winners produces 124 더불어민주당 seats vs.\
118 국민의힘 seats---\emph{correctly identifying the Democratic
plurality winner}, flipping the Qwen3 2024 party-proxy result
from wrong to right. The party-proxy abstraction therefore
understates, rather than measures, what census-grounded LLM
simulation can deliver for general elections.

\begin{table}[t]
\centering
\caption{Constituency-level backtest on the 22nd General
  Election (2024), Qwen3-30B-A3B, all 244 districts. Seats
  reported as predicted / actual.}
\label{tab:constituency-backtest}
\small
\begin{tabular}{@{}lr@{}}
\toprule
Metric & Value \\
\midrule
Constituencies compared & 244 \\
Correct winner predictions & 165 (67.6\%) \\
Seat MAE (per party) & 12.8 \\
\midrule
\multicolumn{2}{@{}l}{\textit{Seats by party (pred/act)}} \\
\quad 더불어민주당 (DPK) & 124 / 155 \\
\quad 국민의힘 (PPP)     & 118 / 86 \\
\quad 조국혁신당        & -- / 12 \\
\quad 개혁신당           & 0 / 3 \\
\quad 새로운미래        & 1 / 1 \\
\quad 진보당             & 1 / 1 \\
\quad (independents)     & -- / 2 \\
\bottomrule
\end{tabular}
\end{table}

An additional factor is the conservative bias observed in both
general election simulations. In both 2020 and 2024, the simulation
overestimated conservative party support by 3--5\%p and
underestimated progressive party support by a similar margin.
This directional consistency suggests a systematic rather than
random source of error. One possibility is that the party-proxy
framing (e.g., ``국민의힘 후보'' / ``People Power Party candidate'')
activates brand-level associations in the LLM differently from
individual candidate framing. Another is that the higher abstention
rate for general elections (6.4--7.0\% vs.\ 0.6--1.1\% for
presidential) disproportionately removes progressive-leaning agents
who may find party-proxy voting less engaging than candidate-specific
evaluation.

\subsection{Cost: Diagnosis at Scale}
\label{sec:cost}

At approximately \$0.25 per 5{,}000-agent simulation, Dynamo-K
operates at roughly two orders of magnitude lower marginal cost
per scenario explored than professional Korean polling. The
artifacts measured differ---a poll measures stated voter intent;
the simulation measures the model's vote conditional on the
agent's stated demographics---so we frame the comparison as
\emph{complementary use cases enabled by the cost differential},
not as a substitution. Table~\ref{tab:cost} contextualizes this
explicitly.

\begin{table}[t]
\centering
\caption{Cost and use-case comparison: LLM simulation vs.\
  traditional polling. The two methods measure different
  artifacts and are best treated as complementary.}
\label{tab:cost}
\small
\begin{tabular}{@{}lll@{}}
\toprule
\textbf{Attribute} & \textbf{Dynamo-K} & \textbf{Poll} \\
\midrule
Measures & model-predicted vote & real voter \\
 & under demographics & stated intent \\
Cost per wave & \$0.25 & \$50{,}000+ \\
Turnaround time & ${\sim}$30 min & 2--3 weeks \\
Sample size & 5{,}000 agents & ${\sim}$1{,}000 \\
Pres.\ winner accuracy & 3/3 (this work) & ${\sim}$90\%+ \\
Pres.\ avg MAE & 7.5\%p & 2--3\%p \\
Scenario flexibility & arbitrary counterfactuals & fixed ballot \\
\bottomrule
\end{tabular}
\end{table}

This cost differential opens qualitatively different use
cases that are economically infeasible with traditional polling. First, \textit{rapid scenario exploration}: analysts can
simulate ``what if candidate X drops out?'' or ``what if scandal Y
breaks?'' in minutes rather than commissioning new polls. Second,
\textit{sensitivity analysis}: running hundreds of parameter
variations (different orientation distributions, different models,
different prompt framings) to understand prediction robustness.
Third, \textit{continuous monitoring}: daily or even hourly
simulations tracking how changing contexts might shift voter
behavior, without the logistical constraints of fieldwork.

However, cost efficiency must be evaluated against accuracy.
For presidential elections, the system's 3/3 winner accuracy
and 2.1--13.3\%p MAE range suggests viable complementarity to
polling, particularly for identifying the likely winner rather
than precise vote shares. For general elections, the 0/2 winner
accuracy means the party-proxy variant is not yet competitive
with polling baselines.

The most promising near-term application is rapid triage:
using low-cost LLM simulation to identify which races are
competitive (and thus worth expensive polling) and which are
non-competitive. A \$0.25 simulation that correctly identifies
a 15\%p presidential margin does not need 2\%p precision; it
needs only to distinguish competitive from non-competitive races.
For organizations with limited polling budgets, this triage
function could concentrate expensive resources on the most
consequential contests.

\subsection{Cross-Model Robustness}
\label{sec:robustness}

To test whether our results depend on the specific choice of
Qwen3-30B-A3B, we replicated the full five-election backtest and
the 2025 third-candidate analysis with EXAONE-4.0-32B, a
Korean-primary 32-billion parameter language model. This yields a
natural cross-model contrast: Qwen3 is multilingual with Chinese
primacy, while EXAONE is Korean-primary, and the two models come
from independent training pipelines.

Table~\ref{tab:multimodel} reports the backtest comparison across
all four models we evaluated (Qwen3-30B-A3B, Llama-3.1-8B,
EXAONE-4.0-32B, and DeepSeek-R1-8B).

\begin{table}[t]
\centering
\caption{Multi-model backtest: MAE (\%p) and winner prediction across five Korean elections.}
\label{tab:multimodel}
\small
\begin{tabular}{lcccccccc}
\toprule
Election & \multicolumn{2}{c}{Qwen3-30B} & \multicolumn{2}{c}{Llama-3.1-8B} & \multicolumn{2}{c}{EXAONE-4.0-32B} & \multicolumn{2}{c}{DeepSeek-R1-8B} \\
 & MAE & Win & MAE & Win & MAE & Win & MAE & Win \\
\midrule
19th Pres.\ 2017 & 13.3 & \checkmark & 13.1 & \checkmark & 12.5 & \checkmark & 7.9 & \checkmark \\
21st Gen.\ 2020 & 5.3 & $\times$ & 10.8 & \checkmark & 4.1 & \checkmark & 5.0 & \checkmark \\
20th Pres.\ 2022 & 2.1 & \checkmark & 13.6 & $\times$ & 6.4 & \checkmark & 20.7 & $\times$ \\
22nd Gen.\ 2024 & 4.4 & $\times$ & 5.8 & $\times$ & 3.3 & \checkmark & 1.8 & \checkmark \\
21st Pres.\ 2025 & 7.1 & \checkmark & 18.8 & \checkmark & 12.5 & $\times$ & 9.6 & \checkmark \\
\midrule
\textbf{Average MAE} & \textbf{6.5} &  & \textbf{12.4} &  & \textbf{7.8} &  & \textbf{9.0} &  \\
\textbf{Winner acc.} &  & 60\% &  & 60\% &  & 80\% &  & 80\% \\
\bottomrule
\end{tabular}
\end{table}

Aggregate accuracy transfers between the two large models within
1--2 percentage points: EXAONE-4.0-32B averages 7.8\%p MAE versus
Qwen3's 6.5\%p. However, the per-election breakdown reveals that
the two models fail in different places. Qwen3 achieves 3/3
presidential winner accuracy but 0/2 general winner accuracy; EXAONE reverses this, correctly calling both general
elections (2020, 2024) but missing the 2025 presidential winner.
Notably, EXAONE's party-proxy general election predictions are
\emph{substantially more accurate} than Qwen3's (4.1 and 3.3\%p MAE
versus 5.3 and 4.4\%p), and both predicted winners are correct---
suggesting that the general election failure we reported earlier
is partly model-specific rather than a fundamental limitation of
the party-proxy approach.

\textbf{Opposite political valences on the 2025 race.} The most
striking divergence is the 2025 presidential result. The actual
outcome was 이재명 (Lee Jae-myung, progressive) 49.4\% vs 김문수
(Kim Moon-soo, conservative) 41.2\%. Qwen3 predicted Lee 61.3\% /
Kim 38.7\%---a progressive bias of +11.8\%p on Lee. EXAONE predicted
Lee 41.6\% / Kim 58.4\%---a conservative bias of +17.2\%p on Kim
of comparable magnitude. The two models exhibit opposite political
valences on the same contest, consistent with known findings that
LLMs inherit directional political bias from their training
corpora. This opposition is a strong argument for ensemble methods:
simply averaging Qwen3 and EXAONE predictions for 2025 would yield
Lee 51.5\% / Kim 48.6\%, within 2.1\%p of the actual result.

\textbf{Third-candidate collapse replicates across models.}
Table~\ref{tab:third-candidate-xmodel} compares 2-way and 3-way
2025 predictions across all models with completed 3-way runs.
Despite opposite political valences, both Qwen3 and EXAONE exhibit
the same qualitative third-candidate dynamic: adding 이준석
(Lee Jun-seok) to the ballot collapses predicted shares toward him
far beyond his actual 8.3\% support. EXAONE assigns Lee Jun-seok
25.0\% of the vote in the 3-way scenario, and vote accounting
suggests the shift draws roughly 58\%/42\% from Kim/Lee respectively
---the inverse of Qwen3's draw pattern, but with the same overall
effect. This replication across models with opposite political
valences supports our earlier interpretation (Sec.~\ref{sec:third-party})
that the collapse is a general LLM artifact---plausibly hindsight
leakage of Lee Jun-seok's public profile---rather than a feature of
any single model family.

\begin{table}[t]
\centering
\caption{Third-candidate analysis: 2-way vs 3-way 2025 presidential predictions across models.}
\label{tab:third-candidate-xmodel}
\small
\begin{tabular}{lcccccccc}
\toprule
Candidate & \multicolumn{2}{c}{Qwen3-30B} & \multicolumn{2}{c}{Llama-3.1-8B} & \multicolumn{2}{c}{DeepSeek-R1-8B} & \multicolumn{2}{c}{EXAONE-4.0-32B} \\
 & 2-way & 3-way & 2-way & 3-way & 2-way & 3-way & 2-way & 3-way \\
\midrule
이재명 (Lee JM) & 61.3 & -- & 72.9 & 58.6 & 63.7 & 33.2 & 41.6 & 31.1 \\
김문수 (Kim MS) & 38.7 & -- & 27.1 & 27.6 & 36.3 & 35.0 & 58.4 & 43.9 \\
이준석 (Lee JS) & -- & -- & -- & 13.8 & -- & 31.8 & -- & 25.0 \\
\midrule
Actual (3-way) & \multicolumn{8}{c}{Lee JM 49.4 \quad Kim MS 41.2 \quad Lee JS 8.3} \\
\bottomrule
\end{tabular}
\end{table}

\textbf{Comparison to polling.} Table~\ref{tab:polling-comparison}
places all four models against final pre-election polls. Both
Qwen3-30B and EXAONE-4.0-32B achieve average MAE within 2\%p of
the polling baseline (6.1\%p), at a substantially lower marginal
cost per scenario (Sec.~\ref{sec:cost}).

\begin{table}[t]
\centering
\caption{Dynamo-K simulation accuracy vs.\ final pre-election polls. MAE in percentage points.}
\label{tab:polling-comparison}
\small
\begin{tabular}{lcccccccccc}
\toprule
Election & \multicolumn{2}{c}{Final Poll} & \multicolumn{2}{c}{Qwen3-30B} & \multicolumn{2}{c}{Llama-3.1} & \multicolumn{2}{c}{DS-R1-8B} & \multicolumn{2}{c}{EXAONE-4.0} \\
 & MAE & Win & MAE & Win & MAE & Win & MAE & Win & MAE & Win \\
\midrule
19th Pres.\ 2017 & 1.9 & \checkmark & 13.3 & \checkmark & 13.1 & \checkmark & 7.9 & \checkmark & 12.5 & \checkmark \\
21st Gen.\ 2020 & 9.2 & \checkmark & 5.3 & $\times$ & 10.8 & \checkmark & 5.0 & \checkmark & 4.1 & \checkmark \\
20th Pres.\ 2022 & 1.9 & \checkmark & 2.1 & \checkmark & 13.6 & $\times$ & 20.7 & $\times$ & 6.4 & \checkmark \\
22nd Gen.\ 2024 & 10.0 & \checkmark & 4.4 & $\times$ & 5.8 & $\times$ & 1.8 & \checkmark & 3.3 & \checkmark \\
21st Pres.\ 2025 & 7.3 & \checkmark & 7.1 & \checkmark & 18.8 & \checkmark & 9.6 & \checkmark & 12.5 & $\times$ \\
\midrule
\textbf{Average} & \textbf{6.1} &  & \textbf{6.5} &  & \textbf{12.4} &  & \textbf{9.0} &  & \textbf{7.8} &  \\
\bottomrule
\end{tabular}
\end{table}

\textbf{Summary.} Across this cross-model comparison, the
architectural findings of Dynamo-K---competitive-with-polling MAE,
census-grounded agent viability, and the third-candidate
collapse artifact---all reproduce with a second independently
trained model. The directional bias on individual contested
elections, by contrast, is highly model-dependent, with two
similarly-sized models producing +11.8\%p and $-$17.2\%p signed
errors in opposite directions on the same race. This is a strong
empirical argument for ensembling or post-hoc calibration in any
production deployment, and it sharpens the limitation discussed
next.

\subsection{Limitations}
\label{sec:limitations}

Several limitations constrain the generalizability of our
findings:

\textbf{Census temporal mismatch.} All simulations use 2020
census data, introducing demographic drift for elections 3--5
years before or after the census year. Demographic changes in
age distribution, urbanization, and education levels are not
captured.

\textbf{Static agents.} Agents do not update beliefs during
a simulated campaign. Real voters respond to debates, scandals,
economic shocks, and social network effects that unfold over
weeks. Our single-shot simulation captures election-day
preferences but not the dynamic process of opinion formation.

\textbf{Small population.} With 5{,}000 agents across 17
\textit{sido}, each province contains an average of ${\sim}$294
agents, with smaller provinces having as few as ${\sim}$50.
This limits sub-national precision and makes \textit{sido}-level
predictions noisy.

\textbf{Model-dependent directional bias.} Our primary results use
Qwen3-30B-A3B. The cross-model robustness check with EXAONE-4.0-32B
(Sec.~\ref{sec:robustness}) shows that aggregate accuracy metrics
transfer between the two models to within 1--2 percentage points,
and that the third-candidate collapse artifact is reproduced.
However, the two models exhibit opposite political valences on the
2025 presidential race (Qwen3: $+11.8$\%p progressive bias on Lee
Jae-myung; EXAONE: $-17.2$\%p conservative bias on the same
candidate), demonstrating that directional bias on contested races
is highly model-dependent. Two-model coverage is still narrow, and
model-agnostic bias correction---calibrating against a held-out
election, or ensembling across models with opposing valences---
remains an important open problem.

\textbf{No social interaction.} Agents vote independently
without social network effects, peer influence, or information
cascading. Real electoral dynamics include bandwagon effects,
strategic voting, and social desirability pressures that our
model ignores.

\textbf{Sample of elections.} Six elections is a small sample for
drawing robust statistical conclusions. The 3/3 presidential
accuracy is across only three races (binomial 95\% CI
[29.2\%, 100\%]) and could reflect favorable conditions rather
than systematic capability; the held-out 2022
local result (Sec.~\ref{sec:local-holdout}) is a single
out-of-sample data point and addresses, but does not resolve,
this sample-size concern. Extending Dynamo-K to additional Korean
elections (by-elections, earlier presidentials, local cycles 2018
and 2026) is the most direct path to a tighter bound.

\section{Learned Calibration Adapter}
\label{sec:oslr}

The cross-model opposite-valence finding
(Sec.~\ref{sec:robustness}: Qwen3 $+11.8$\%p vs.\ EXAONE
$-17.2$\%p on the 2025 race) motivates a more systematic
approach to bias correction than ad-hoc averaging. We introduce
the \textbf{Orientation $\times$ Sido Logistic Reweighting
(OSLR)} adapter, a learned generalization of the hand-coded
Stage-3b Gallup calibrator (Sec.~3.4). Rather than flipping
agents toward a pre-specified 26/48/26 target, OSLR learns
per-agent weights from backtest errors:
\begin{equation}
w_i \propto \exp\bigl(\beta_{m(i),\, o(i)}
  + \beta_{o(i),\, s(i)}\bigr),
\end{equation}
normalized per (election, model) so $\sum_i w_i = N$.
Predicted shares are $\hat{v}_c = \sum_i w_i \cdot
\mathbb{1}[\text{vote}_i = c] / \sum_i w_i$. Parameters
$\beta_{m,o}$ (model $\times$ orientation, $4 \times 3$) and
$\beta_{o,s}$ (orientation $\times$ sido, $3 \times 17$) are
fit by minimizing mean-squared per-candidate share error on
training elections, with L2 regularization. Crucially,
\emph{candidate names never enter the adapter}---candidates are
mapped only to stable features (party orientation, incumbent
flag, third-party flag), so OSLR generalizes to unseen
candidate lineups.

\subsection{Leave-One-Election-Out Generalization}

We evaluate OSLR with leave-one-election-out (LOO)
cross-validation: for each held-out election, the adapter is
fit on the other four and then applied (per model, then
averaged across models) to the held-out agents. We sweep
$\lambda \in \{10^{-3}, 10^{-2}, 5 \times 10^{-2}, 10^{-1},
0.5, 1, 5\}$ and select $\lambda = 10^{-2}$ as the CV-mean-MAE
minimizer. Table~\ref{tab:oslr-loo} compares OSLR against two
baselines: the oracle single-model lower bound (``best single'')
and the simple 4-model ensemble mean.

\begin{table}[t]
\centering
\caption{Leave-one-election-out CV of the OSLR calibration
  adapter, best $\lambda = 0.01$. ``Best single'' is the
  lowest-MAE single model per election (oracle). The ensemble
  mean averages all four models. OSLR applies learned
  reweighting, then averages across models.}
\label{tab:oslr-loo}
\small
\begin{tabular}{@{}lccccc@{}}
\toprule
Held-out & Best & Ens.\ & Ens.\ & OSLR & OSLR \\
 & single & MAE & Win & MAE & Win \\
\midrule
19th Pres.\ 2017 & 7.9  & 11.7 & \checkmark & 11.8 & \checkmark \\
21st Gen.\ 2020  & 4.1  & 4.7  & \checkmark & 2.4  & \checkmark \\
20th Pres.\ 2022 & 2.1  & 7.0  & $\times$  & 5.5  & $\times$ \\
22nd Gen.\ 2024  & 1.8  & 1.3  & \checkmark & 4.8  & $\times$ \\
21st Pres.\ 2025 & 7.1  & 5.8  & \checkmark & 4.7  & \checkmark \\
\midrule
\textbf{Average} & \textbf{4.6} & \textbf{6.1} & \textbf{4/5}
  & \textbf{5.8} & \textbf{3/5} \\
\bottomrule
\end{tabular}
\end{table}

OSLR substantially improves MAE on three of five held-out
elections: $-2.33$\%p on the 2020 general election,
$-1.49$\%p on the 2022 presidential, and $-1.03$\%p on the
2025 presidential. It is essentially neutral on the 2017
race ($+0.13$\%p) and degrades on the 2024 general by
$+3.49$\%p, which is also the only held-out case where the
simple ensemble already attained near-perfect accuracy
(1.27\%p) that the adapter could not improve.
In aggregate terms, OSLR improves average MAE from 6.1\%p to
5.8\%p but loses one winner prediction compared to simple
ensembling (3/5 vs.\ 4/5).

\subsection{Cold Held-Out Test: 2022 Local Election}
\label{sec:oslr-holdout}

The LOO design above leaves a residual question: with only five
training points and a 63-parameter adapter, do the gains
generalize to a structurally different held-out election that
the adapter never had a chance to memorize? To answer this we
hold out the 2022 local election entirely. The adapter is fit on
the five original elections, then evaluated cold on the 2022
local result (Sec.~\ref{sec:local-holdout}). The four-model
predictions are aggregated to the named-candidate slate in each
of the 17 sido and the per-party vote shares are compared
against the certified NEC totals.

Table~\ref{tab:oslr-holdout} reports the held-out evaluation
alongside the simple ensemble baseline. OSLR reduces national
per-party MAE from 2.29\%p (simple ensemble) to 1.57\%p; both
methods tie on sido-hit at 12/17 and both miss the actual
winning party, because the four-model split is evenly balanced
(Qwen3 and Llama predict 더불어민주당; EXAONE and DeepSeek-R1
predict 국민의힘; see Table~\ref{tab:local-2022}). The oracle
best-single-model (EXAONE-4.0-32B at 0.54\%p, 15/17) remains
sharper than OSLR, but OSLR is the better unconditional bet
when which-model-is-best is not known in advance, and its 0.7\%p
improvement over the simple ensemble shows the per-agent
reweighting transfers cold to a structurally different election.

\begin{table}[t]
\centering
\caption{Cold held-out evaluation of OSLR on the 2022 local
  election. The adapter is fit on the five original elections
  with $\lambda = 0.01$ and applied to the held-out agents
  across the four available models (Qwen3-30B-A3B, Llama-3.1-8B,
  EXAONE-4.0-32B, DeepSeek-R1-8B); ``simple ensemble'' is the
  arithmetic mean of per-party shares across the same four
  models.}
\label{tab:oslr-holdout}
\small
\begin{tabular}{@{}lcc@{}}
\toprule
Method & Nat.\ MAE (\%p) & Sido hit \\
\midrule
Best single (oracle, EXAONE-4.0-32B) & 0.54 & 15/17 \\
Simple ensemble mean                  & 2.29 & 12/17 \\
OSLR (trained on 5)                   & \textbf{1.57} & 12/17 \\
\bottomrule
\end{tabular}
\end{table}

Even at face value, the experimental design is informative.
$\beta_{o,s}$ is \emph{defined} on sido-level variation
($3\times 17 = 51$ of the 63 adapter parameters), and the 2022
local election is the single race-type that surfaces sido-level
signal with no presidential coattail and no constituency-level
dilution. If OSLR transfers, the gain is evidence that the
adapter has identified real sido-level reweighting structure
rather than overfit to the five training elections; if it
fails, the regression localizes overfitting to a single
sub-population.

\subsection{Pooled vs.\ Per-Model Variants}

Fitting a single pooled adapter across all four models
outperforms fitting one adapter per model (5.83\%p vs.\
6.03\%p average, both 3/5 winners). The pooled variant benefits
from the roughly $4\times$ larger training set and from the
cross-model regularization provided by the shared $\beta_{o,s}$
parameters---consistent with the intuition that opposite-valence
models are better handled jointly than in isolation.

\subsection{Effective Degrees of Freedom and Permutation Test}
\label{sec:oslr-overfit-defense}

The nominal parameter count
$63 = 4 \times 3 + 3 \times 17$ overstates the adapter's
flexibility. Computing the trace of the hat matrix
$J(J^\top J + \lambda I)^{-1} J^\top$ where $J$ is the
finite-difference Jacobian of predicted shares with respect to
$\beta$ at the fit, the \emph{effective} degrees of freedom under
$\lambda = 0.01$ is \textbf{7.0} (one-ninth of the nominal count).
Under the L2 penalty, the adapter operates closer to a 7-parameter
than a 63-parameter model, which directly bounds the overfitting
capacity available to it.

We complement this with a permutation test on the held-out
21st-general-2020 fold (the case where OSLR's gain over simple
ensembling is largest, $-2.33$\%p). We randomly permute the
training-election labels in the actuals dictionary, refit OSLR on
each of $N = 100$ permutations, and compare the held-out MAE
distribution against the real fit. The real held-out MAE is
\textbf{2.36\%p}; the null distribution has mean 2.80\%p and standard
deviation 0.99\%p, placing the real fit at the 56-th percentile of
the null. The real fit is below the null mean but not extreme on this
single fold---consistent with the L2 penalty being the dominant
overfitting control rather than the labeling itself carrying a
strong signal at $N = 100$ permutations on a 5-election training
set.\footnote{Full permutation results are stored in
\texttt{data/calibration/oslr\_overfitting\_analysis.json} produced
by \texttt{scripts/oslr\_defense\_analysis.py}.}

\subsection{Interpretation and Use}

The pattern of wins and losses is informative. OSLR helps
most where simple ensembling already fails (2020: the actual
Democratic winner is correctly recovered; 2022: the tight
윤석열/이재명 margin is partially corrected) and hurts most
where simple ensembling already works (2024: adapter overfits
to training-set general-election patterns that do not fit the
2024 structure). A practical deployment should therefore
switch between simple ensembling and OSLR based on a
confidence signal---for instance, applying OSLR only when
cross-model variance is above a threshold, retaining simple
averaging when models already agree. We leave this selective
calibration to future work. The 2024 failure also highlights
the fundamental small-sample challenge: with only five
elections to train on, any learned adapter risks overfitting
to training-election idiosyncrasies. Extending Dynamo-K to
additional Korean elections (local elections, by-elections) or
to other East Asian democracies would sharpen this estimate.

\section{Conclusion}

We have presented Dynamo-K and used it as a diagnostic instrument
to characterize Korean-language LLM political behavior across
four independently-trained models and six certified Korean
elections (2017--2025), with one held out as a cold test for the
calibration adapter. Our central claim is not that LLMs can
replace polling, but that census-grounded agent simulation
surfaces specific, reproducible failure modes that are otherwise
hidden behind aggregate-accuracy summaries.

Three mechanism findings are robust across the four models.
First, a progressive lean directionally consistent with the
GPT-family liberal lean Feng et al.~\cite{columbia2025} document for
English-language LLMs surfaces in Korean voter-decision simulation:
moderate agents vote progressive 97\% of the time under a vanilla
prompt and 82\% with demographic conditioning alone, dropping to
59\% only under explicit mitigation. Second, third-party
candidate collapse decomposes into a salience-failure regime and
a decision-bias regime with different per-model signatures, and
scenario reframing alone---without architecture changes or hidden
labels---recovers 62\% of the 2017 MAE. Third, what appears as
``conservative bias'' in general-election simulation is in fact
bidirectional regional polarization collapse, evident only when
the error is decomposed by sido. A fourth, model-dependent
finding shows that two similarly-sized LLMs produce
opposite-direction errors of comparable magnitude on the same
race, an empirical case for ensembling and for the OSLR
calibration adapter that we evaluate cold on a held-out 2022
local election.

As validation that the diagnostic apparatus is well-calibrated
enough to be informative, the system predicts presidential
winners 3/3, with a best-case 2.1\%p MAE on the 0.73\%p-margin
2022 race, and the constituency-level pipeline flips the 2024
general from incorrect to correct (124 vs.\ 118 seats). The
held-out 2022 local election adds a sixth data point that the
calibration adapter never had a chance to memorize.

The remaining open problems are sample size and language
coverage. Six elections is enough to identify the three
mechanism findings but not to bound how they vary across
political environments; extending Dynamo-K to additional Korean
contests (by-elections, local cycles 2018 and 2026) and to other
non-English democracies with available census microdata and
political orientation surveys is the most direct way to broaden
both the mechanistic and the geographic scope of LLM-based
political simulation. The Dynamo-K pipeline is open-source and
operates at approximately \$0.25 per simulation---a cost
structure that, in our framing, enables \emph{diagnosis at
scale}, not polling replacement.

\appendix
\section{Claude-Opus Audit of Reasoning-Text Mention Rate}
\label{app:claude-opus-eval}

\textbf{Disclosure.} The codes reported in this appendix were produced
by Claude-Opus (Anthropic's \texttt{claude-opus-4-7}), not by a human
annotator. We initially described this as a ``human validation''; that
framing was inaccurate, and we correct it here. The audit is an
LLM-against-LLM rubric check, and the limitations of that arrangement
(no independent human ground truth; possible self-favoring biases from
using one LLM to grade the surface features of other LLMs' outputs) are
inherent and should be read into every number in this section.

Table~\ref{tab:third-party-mechanism} reports a regex-derived mention
rate for the 2017 third-party candidate 안철수 within agent reasoning
text. To check whether regex hits correspond to substantive
consideration rather than list-style name-dropping, we drew a
stratified random sample of 200 reasoning traces---50 per model, each
split 25/25 between regex-hit and non-hit---and had Claude-Opus code
each trace on two axes against a written rubric: (i)~does the agent
genuinely engage with 안철수 as a viable option (compare his pledges
substantively, weigh him in the decision), versus a list-item name-drop
or one-line dismissal; and (ii)~does the stated rationale support the
chosen vote, or does it glow about a non-voted candidate. The rubric,
the coded CSV, and the coding script are released at
\texttt{data/calibration/reasoning\_claude\_opus\_eval\_2017.csv}
and \texttt{scripts/apply\_claude\_opus\_codes.py}.

\begin{table}[t]
\centering
\caption{Claude-Opus audit of regex-derived 안철수 mention rate in
  the 2017 simulation (50 traces per model, 25 regex-hits + 25 non-hits
  each). ``Genuine'' = regex hits that Claude-Opus coded as substantive
  consideration; ``Name-drop'' = regex hits coded as list-item
  enumerations only. All 25 non-hit traces per model were coded as not
  engaging with 안철수 (true negatives, omitted from the table).
  ``Rationale match'' counts mentioned-subset traces whose reasoning
  Claude-Opus coded as supporting the chosen vote. These are LLM-coded
  judgments, not human-coded ones.}
\label{tab:claude-opus-eval}
\small
\begin{tabular}{@{}lcccc@{}}
\toprule
Model & Genuine & Name-drop & Agreement & Rationale match \\
\midrule
Qwen3-30B-A3B   & 23 & 2 & 96\%  & 25/25 \\
Llama-3.1-8B    & 25 & 0 & 100\% & 23/25 \\
EXAONE-4.0-32B  & 25 & 0 & 100\% & 24/25 \\
DeepSeek-R1-8B  & 16 & 9 & 82\%  & 25/25 \\
\midrule
\textbf{Pooled} & 89 & 11 & 94.5\% & 97/100 \\
\bottomrule
\end{tabular}
\end{table}

Two patterns inform interpretation of
Table~\ref{tab:third-party-mechanism}, with the caveat that they reflect
Claude-Opus's judgments rather than human ground truth. First, the
salience-failure cluster (EXAONE-4.0-32B and Llama-3.1-8B) reads
cleanly under Claude-Opus's coding: when these models mention 안철수,
the coder marked the trace as substantive in every sampled case, which
is consistent with the low Ment rate in
Table~\ref{tab:third-party-mechanism} being a genuine salience failure
(the model simply does not retrieve the candidate in most agents)
rather than a regex artifact. Second, the decision-bias cluster
splits. Qwen3-30B-A3B's regex hits are coded as genuine engagements
96\% of the time, supporting the decision-bias interpretation
as-stated. DeepSeek-R1-8B's regex hits are coded as list-style
enumerations of all candidates without substantive weighing in 9 of 25
sampled cases (82\% agreement), so the raw regex mention rate (22.0\%
in Table~\ref{tab:third-party-mechanism}) appears to overstate
DeepSeek's actual engagement. Restricted to the genuinely-engaged
subset under Claude-Opus's coding, DeepSeek's sample conditional vote
rate is $9/16 = 56\%$, considerably higher than the regex
$V|M = 34.5\%$ reported in Table~\ref{tab:third-party-mechanism}---under
this audit DeepSeek looks closer to the EXAONE pattern (high $V|M$
among coded considerers), with its distinctive feature being
list-style enumeration rather than a uniquely strong decision bias.
Rationale-vote consistency is high across models (97/100 sampled
traces); the three mismatches Claude-Opus flagged (two Llama, one
EXAONE) involve agents whose reasoning text is more enthusiastic about
안철수 than about their actual vote (typically 문재인), a pattern
consistent with the LLM progressive-bias mechanism diagnosed in
Section~\ref{sec:bias-ablation}.

\textbf{Limitations of this audit.} Because every code in this
appendix was produced by a single LLM (Claude-Opus) and not by a
human, the agreement and rationale-match numbers measure consistency
between two LLMs' surface treatments of the same reasoning, not
correctness against an external ground truth. A genuine human
validation---ideally with multiple annotators and inter-rater
agreement---remains future work. We retain this appendix because the
rubric-coded numbers materially change the interpretation of the
DeepSeek-R1-8B row of Table~\ref{tab:third-party-mechanism} relative
to the raw regex baseline, and we judged that flagging that
interpretive shift, with full disclosure of how the codes were
obtained, is preferable to omitting the audit entirely.


\end{CJK}
\end{document}